\documentclass[11pt]{article}

\renewcommand{\theequation}{\arabic{section}-\arabic{equation}}
\newcommand{\stln}{\setlength{\unitlength}{2.2ex}}
\newcommand{\fr}{\framebox(1,1){}}
\newcommand{\sfr}{\framebox(1,1){\begin{picture}(1,1)
  \put(0,0){\line(1,1){1}}
\end{picture}}}

\newcommand{\onebox}
{\stln \lower1.4ex\hbox{
\begin{picture}(1.6,1.6)
\put(.3,.3){\fr}
\end{picture}}}
\newcommand{\twobox}
{\stln \lower1.4ex\hbox{
\begin{picture}(2.6,1.6)
\put(.3,.3){\fr}
\put(1.3,.3){\fr}
\end{picture}}}
\newcommand{\threebox}
{\stln \lower1.4ex\hbox{
\begin{picture}(3.6,1.6)
\multiput(.3,.3)(1,0){3}{\fr}
\end{picture}}}
\newcommand{\fourbox}
{\stln \lower1.4ex\hbox{
\begin{picture}(4.6,1.6)
\multiput(.3,.3)(1,0){4}{\fr}
\end{picture}}}
\newcommand{\fivebox}
{\stln \lower1.4ex\hbox{
\begin{picture}(5.6,1.6)
\multiput(.3,.3)(1,0){5}{\fr}
\end{picture}}}
\newcommand{\oneonebox}
{\stln \lower2.6ex\hbox{
\begin{picture}(1.6,2.6)
\put(.3,.3){\fr}
\put(.3,1.3){\fr}
\end{picture}}}
\newcommand{\twoonebox}
{\stln \lower2.6ex\hbox{
\begin{picture}(2.6,2.6)
\put(.3,1.3){\fr}
\put(1.3,1.3){\fr}
\put(0.3,0.3){\fr}
\end{picture}}}
\newcommand{\twotwobox}
{\stln \lower2.6ex\hbox{
\begin{picture}(2.6,2.6)
\put(.3,.3){\fr}
\put(.3,1.3){\fr}
\put(1.3,.3){\fr}
\put(1.3,1.3){\fr}
\end{picture}}}
\newcommand{\marcnbox}
{\stln \lower1.4ex \hbox{
\begin{picture} (6.6, 3.1)
\multiput(.3, .3) (1, 0) {2} {\fr}
\put(2.3, .3) {\framebox(3,1){$\cdots$}}
\put(5.3, .3){\fr}
\put(.3, 1.4)
{$\overbrace{~~~~~~~~~~~~~~~~~~}^{n}$}
\end{picture}}}
\newcommand{\marcnplusonebox}
{\stln \lower1.4ex \hbox{
\begin{picture} (6.6, 3.1)
\multiput(.3, .3) (1, 0) {2} {\fr}
\put(2.3, .3) {\framebox(3,1){$\cdots$}}
\put(5.3, .3){\fr}
\put(.3, 1.4)
{$\overbrace{~~~~~~~~~~~~~~~~~~}^{n+1}$}
\end{picture}}}
\newcommand{\marcnplustwobox}
{\stln \lower1.4ex \hbox{
\begin{picture} (6.6, 3.1)
\multiput(.3, .3) (1, 0) {2} {\fr}
\put(2.3, .3) {\framebox(3,1){$\cdots$}}
\put(5.3, .3){\fr}
\put(.3, 1.4)
{$\overbrace{~~~~~~~~~~~~~~~~~~}^{n+2}$}
\end{picture}}}
\newcommand{\marcnplusthreebox}
{\stln \lower1.4ex \hbox{
\begin{picture} (6.6, 3.1)
\multiput(.3, .3) (1, 0) {2} {\fr}
\put(2.3, .3) {\framebox(3,1){$\cdots$}}
\put(5.3, .3){\fr}
\put(.3, 1.4)
{$\overbrace{~~~~~~~~~~~~~~~~~~}^{n+3}$}
\end{picture}}}
\newcommand{\marcnplusfourbox}
{\stln \lower1.4ex \hbox{
\begin{picture} (6.6, 3.1)
\multiput(.3, .3) (1, 0) {2} {\fr}
\put(2.3, .3) {\framebox(3,1){$\cdots$}}
\put(5.3, .3){\fr}
\put(.3, 1.4)
{$\overbrace{~~~~~~~~~~~~~~~~~~}^{n+4}$}
\end{picture}}}
\newcommand{\sonebox}
{\stln \lower1.4ex\hbox{
\begin{picture}(1.6,1.6)
\put(.3,.3){\sfr}
\end{picture}}}
\newcommand{\smarctwojbox}
{\stln \lower1.4ex \hbox{
\begin{picture}(6.6,3.1)
\multiput(.3,.3)(1,0){2}{\sfr}
\put(2.3,.3){\framebox(3,1){$\cdots$}}
\put(5.3,.3){\sfr}
\put(.3,1.4){$\overbrace{~~~~~~~~~~~~~~~~~~}^{2j}$}
\end{picture}}}

\newcommand{\OS}{$OSp(8^{*}|4)$}
\newcommand{\OSN}{$OSp(8^{*}|2N)$}
\newcommand{\eq}{\begin{equation}}
\newcommand{\en}{\end{equation}}
\newcommand{\eqn}{\begin{eqnarray}}
\newcommand{\enn}{\end{eqnarray}}
\newcommand{\nn}{\nonumber}

\begin{document}

\begin{titlepage}
\begin{center}
{\bf SUPERCOHERENT STATES OF $OSp(8^*|2N)$, CONFORMAL SUPERFIELDS AND THE $AdS_{7}/CFT_{6}$
 DUALITY }\footnote{Work supported in part by the National Science Foundation under Grant Number
 PHY-9802510} \\
\vspace{1cm}
{\bf Sudarshan Fernando}\footnote{e-mail: fernando@phys.psu.edu } \\
{\bf Murat G\"{u}naydin}\footnote{e-mail: murat@phys.psu.edu } \\
{\bf Seiji Takemae}\footnote{e-mail: takemae@phys.psu.edu.} \\
\vspace{.4cm}
Physics Department \\ Penn State University \\
University Park, PA 16802 \\
\vspace{.5cm}
{\bf Abstract}
\end{center}
We study the positive energy unitary representations of $2N$ extended superconformal
algebras \OSN ~in six dimensions. These representations can be formulated in a particle
basis or a supercoherent state basis, which are labeled by the superspace coordinates in
$d=6$. We show that the supercoherent states that form the bases of  positive energy
representations of \OSN ~can be identified with conformal superfields in six dimensions.
  The massless conformal superfields correspond precisely to the ultra short doubleton
  supermultiplets of \OSN. The other positive energy unitary representations correspond to
  massive conformal superfields in six dimensions and they can be obtained by tensoring an
  arbitrary number of doubleton supermultiplets with each other. The supermultiplets obtained
  by tensoring two copies of the doubletons correspond to massless anti-de Sitter supermultiplets
  in $d = 7$.
\end{titlepage}

\renewcommand{\theequation}{\arabic{section} - \arabic{equation}}
\section{Introduction}
\setcounter{equation}{0} $AdS/CFT$ dualities in M/superstring
theory have been studied intensively over the  last several years.
These dualities relate M/superstring theory over the product
spaces of $d$-dimensional $AdS$ spaces with compact Einstein
manifolds to large $\mathcal{N}$ limits of certain conformal field
theories in $(d-1)$-dimensions. The conjecture of Maldacena that
started the recent interest on $AdS/CFT$ dualities \cite{mald} was
formulated in a more precise manner in \cite{pol,witt}. It
represents the culmination of earlier work on the physics of $N$
$Dp$-branes in the near horizon limit \cite{ads} and much earlier
work on the construction of the Kaluza-Klein spectra of IIB and
eleven dimensional supergravity theories \cite{mgnm}-\cite{ss}.
The relation of Maldacena's conjecture to earlier work on
Kaluza-Klein supergravity theories has been studied in
\cite{mgdm,dupo}. For an extensive review of $AdS/CFT$ dualities
and the references on the subject, we refer the reader to
\cite{adscft}.

 In this paper, we study the unitary supermultiplets
of \OSN. Our work represents an extension of earlier studies of
the representations of this supergroup \cite{gnw,mgst}. For $N =
1,2$ these supermultiplets are important in the study of
$AdS_{7}/CFT_{6}$ dualities in M/superstring theory. Our main
focus is the formulation of the positive energy unitary
representations of \OSN ~in a non-compact supercoherent state
basis. We show that these non-compact supercoherent states
correspond to conformal superfields in $d = 6$. Of these, the
massless conformal superfields are described by the ultra short
doubleton supermultiplets and the massive conformal superfields
are obtained by tensoring an arbitrary number of these doubletons.

The non-compact supergroup \OSN ~can also be interpreted as the
$2N$ extended $AdS$ supergroup in $d=7$. In fact, the symmetry
superalgebra of M-theory compactified to $AdS_{7}$ over $S^{4}$ is
\OS ~\cite{ptn,gnw}. The general method for the oscillator
construction of unitary supermultiplets of \OS ~was first given in
\cite{gnw} with emphasis on short supermultiplets that appear in
the Kaluza-Klein compactification of eleven dimensional
supergravity theory. The entire Kaluza-Klein spectrum of the
eleven dimensional supergravity over $AdS_7 \times S^4$ can be
obtained by a simple tensoring procedure from the ``CPT
self-conjugate'' doubleton supermultiplet \cite{gnw}. This ``CPT
self-conjugate'' doubleton supermultiplet is simply the $(2,0)$
conformal supermultiplet of the dual field theory in six
dimensions. The doubleton supermultiplets of \OSN ~do not have a
Poincar\'{e} limit in $d=7$. By tensoring two copies of these
doubletons, one can obtain all the  massless supermultiplets of
$2N$ extended $AdS$ superalgebra in $d=7$ \cite{mgst}. Tensoring
more than two copies leads to massive $AdS_{7}$ supermultiplets.
The $AdS_7/CFT_6$ duality has been studied from various points of
view more recently \cite{recent,fs}.

 More specifically, in Section
2 we discuss the coherent states associated with positive energy
unitary representations of $SO^{*}(8)$ and show that they
correspond to conformal fields in $d = 6$. Of particular interest
are the doubleton representations of $SO^{*}(8)$, which correspond
to massless conformal fields in six dimensions.

 In Section 3, we
discuss the compact versus non-compact bases of the supergroup
\OSN. In the compact basis, we work with superoscillators, which
transform covariantly under the maximal compact subsupergroup
$U(4|N)$ of \OSN. On the other hand, in the non-compact basis, we
work with operators which transform covariantly under $SU^{*}(4)
\times USp(2N)$ and that have a definite conformal dimension.

 In Section 4 we show how to define a supercoherent state basis for
each positive energy unitary irreducible representation of \OSN.
These supercoherent states correspond to conformal superfields in
six dimensions. As mentioned above, the doubleton supercoherent
states lead to massless conformal superfields, a complete list of
which is given in Section 4.1. By tensoring doubletons with each
other, one obtains massive conformal supermultiplets, of which
those that are obtained by tensoring two copies of doubletons
correspond to massless supermultiplets in $AdS_{7}$ space. For $N
= 2$, the shortest such supermultiplet is the massless $AdS$
graviton supermultiplet of \OS. We give the explicit expression
for the corresponding supercoherent state in Section 4.2.

 We conclude with a discussion of our results.

\section{Coherent states of the positive energy unitary representations of the group
$SO^{*}(8)$ and conformal fields in six dimensions}
\setcounter{equation}{0}
The commutation relations of the generators of the conformal group $SO(6,2)$ in $d = 6$,
which is isomorphic to $SO^{*}(8)$, can be written as
\eqn
\left[ M_{\mu \nu} , M_{\rho \sigma} \right] & = & i ( \eta_{\nu \rho}
 M_{\mu \sigma} - \eta_{\mu \rho} M_{\nu \sigma} - \eta_{\nu \sigma}
 M_{\mu \rho} + \eta_{\mu \sigma} M_{\nu \rho} ), \nn \\
\left[ P_{\mu} , M_{\rho \sigma} \right] & = & i ( \eta_{\mu \rho} P_{\sigma}
 - \eta_{\mu \sigma} P_{\rho} ), \nn \\
\left[ K_{\mu} , M_{\rho \sigma} \right] & = & i ( \eta_{\mu \rho}
K_{\sigma} - \eta_{\mu \sigma} K_{\rho} ), \nn \\
\left[ D , M_{\mu \nu} \right] & = & \left[ P_{\mu} , P_{\nu} \right] \hspace{0.2cm}
= \hspace{0.2cm} \left[ K_{\mu} , K_{\nu} \right] \hspace{0.2cm} = \hspace{0.2cm} 0, \nn \\
\left[ P_{\mu} , D \right] & = & i P_{\mu}, \nn \\
\left[ K_{\mu} , D \right] & = & - i K_{\mu}, \nn \\
\left[ P_{\mu} , K_{\nu} \right] & = & 2 i ( \eta_{\mu \nu} D - M_{\mu \nu} ),
\enn
where $M_{\mu \nu}$ are the generators of the Lorentz subgroup $SO(5,1)$, $D$ is the
 dilatation generator and $P_{\mu}$ and $K_{\mu}$ are the generators of translations and
 special conformal transformations, respectively ($\mu,\nu,\rho,\sigma = 0,1,\ldots,5$)
 \footnote{A complete list of indices we used in this paper is given in Appendix A.}.
 We use the Minkowski metric $\eta_{\mu \nu} = diag (+,-,-,-,-,-)$.
 The covering group of the conformal group $SO(6,2)$ is $Spin(6,2)$ and the
  covering group of the Lorentz group $SO(5,1)$ is $SU^*(4)$. The rotation subgroup $SO(5)$
   (or its covering group $USp(4)$) is generated by $M_{\mu \nu}$ with $\mu,\nu = 1,2,\ldots,5$.

The isomorphism of the conformal group to $SO^{*}(8)$ becomes manifest by defining
\eq
M_{\mu 6} := \frac{1}{2} (P_{\mu} - K_{\mu}), \quad M_{\mu 7} := \frac{1}{2} (P_{\mu}
 + K_{\mu}), \quad M_{67} := -D,
\en
as one finds that together with $M_{\mu \nu}$ they satisfy
\eq
\left[ M_{ab} , M_{cd} \right] = i ( \eta_{bc} M_{ad} - \eta_{ac} M_{bd} - \eta_{bd} M_{ac}
+ \eta_{ad} M_{bc} ), \label{SO42}
\en
where $- \eta_{66} = \eta_{77} = 1$ and $a,b,c,d = 0,1,\ldots,7$.

Considered as the isometry group of the seven dimensional anti-de Sitter space $AdS_7$,
the generators of $SO^{*}(8)$ acquire a different physical interpretation. In particular,
the rotation group becomes $SO(6)$, with the covering group $SU(4)$, generated by
$M_{mn}$ ($m,n = 1,2,\ldots,6$). The generator $E \equiv M_{07}$ becomes the $AdS$ energy,
generating translations along the timelike Killing vector field of $AdS_{7}$ and together
with $M_{mn}$, it generates the maximal compact subgroup $U(4) = SU(4) \times U(1)_{E}$ of
$SO^{*}(8)$.

The Lie algebra of the conformal group $SO(6,2)$, $g$ has a 3-graded decomposition with
respect to its maximal compact subalgebra $L^{0} = SU(4) \times U(1)_{E}$ :
\eq
g = L^{-} \oplus L^{0} \oplus L^{+},
\en
where
\eqn
\left[ L^{0} , L^{\pm} \right] & \subseteq & L^{\pm}, \nn \\
\left[ L^{+} , L^{-} \right] & \subseteq & L^{0}, \nn \\
\left[ L^{+} , L^{+} \right] & = & \left[ L^{-} , L^{-} \right] \hspace{0.2cm} = \hspace{0.2cm}
0, \nn \\
\left[ E , L^{\pm} \right] & = & \pm L^{\pm}, \quad \left[ E , L^{0} \right] \hspace{0.2cm}
 = \hspace{0.2cm} 0.
\enn

The 3-grading is determined by the $U(1)_{E}$ generator $ E = \frac{1}{2} ( P_{0} + K_{0} )$,
which is simply the conformal Hamiltonian. In the oscillator construction of the positive
energy unitary representations of $SO^{*}(8)$, one first realizes its generators as bilinears
 of an arbitrary number $P$ (``generations'' or ``colors'') of pairs of bosonic annihilation
 (${\bf a}_{i}, {\bf b}_{j}$) and creation (${\bf a}^{i}, {\bf b}^{j}$) operators
 ($i,j = 1,2,3,4$), transforming in the fundamental representation of $SU(4)$ and its
 conjugate, respectively \cite{mgcs,ibmg,gnw,mgrs,mgst} :
\eqn
A_{ij} & := & {\bf a}_{i} \cdot {\bf b}_{j} - {\bf a}_{j} \cdot {\bf b}_{i}, \nn \\
A^{ij} & := & {\bf a}^{i} \cdot {\bf b}^{j} - {\bf a}^{j} \cdot {\bf b}^{i}, \nn \\
M^{i}_{~j} & := & {\bf a}^{i} \cdot {\bf a}_{j} + {\bf b}_{j} \cdot {\bf b}^{i}, \label{CB}
\enn
where ${\bf a}_{i} \cdot {\bf b}_{j} := \sum_{r=1}^{P} a_i(r) b_j(r)$, etc. The bosonic
 annihilation and creation operators $a^{i}(r) = a_{i}(r)^{\dag}$ and $b^{j}(r) =
 b_{j}(r)^{\dag}$ satisfy the usual canonical commutation relations
\eqn
\left[ a_{i}(r) , a^{j}(s) \right] & = & \delta_{i}^{~j} \delta_{rs}, \nn \\
\left[ b_{i}(r) , b^{j}(s) \right] & = & \delta_{i}^{~j} \delta_{rs},
\enn
where $i,j = 1,2,3,4$ and $r,s = 1,2,\ldots,P$.

$M^{i}_{~j}$ are the generators of the maximal compact subgroup $U(4)$. The trace part,
 $M^{i}_{~i}$ generates the $AdS$ energy given by
\eqn
Q_{B} & := & \frac{1}{2} M^{i}_{~i} \nn \\
& = & \frac{1}{2} \left( N_{B} + 4 P \right),
\enn
where $N_{B} \equiv {\bf a}^{i} \cdot {\bf a}_{i} + {\bf b}^{i} \cdot {\bf b}_{i}$,
which is the bosonic number operator. The energy eigenvalues of $Q_{B}$ are denoted as $E$.

The hermitian linear combinations of $A_{ij}$ and $A^{ij}$ are the non-compact generators of
 $SO(6,2)$ \cite{mgcs,gnw,mgst}.

Practically in all applications to fundamental physics, the relevant representations of
the conformal group ($AdS$ group) are the unitary irreducible representations (UIRs) of
the lowest weight type, in which the spectrum of the conformal Hamiltonian (the $AdS$
energy)  $E$ is bounded from below. The natural basis for constructing them is the
compact basis in which the lowest weight (positive energy) property as well as unitarity
are manifest.

The lowest weight UIRs of $SO(6,2)$ can  be constructed in a simple way by
using the oscillator realization of the generators given above. Each lowest weight UIR is
uniquely determined by the quantum numbers of a lowest weight vector $| \Omega \rangle$,
provided that $| \Omega \rangle$ transforms irreducibly under $SU(4) \times U(1)_{E}$ and
is also annihilated by all the elements of $L^{-}$ \cite{gnw,mgst}.

A complete list of
possible lowest weight vectors for $P = 1$, which are called doubleton representations in
this compact basis is \cite{mgst}, \eqn
| 0 \rangle, & & \nn \\
a^{i_1} | 0 \rangle & = & | \onebox \rangle, \nn \\
a^{( i_1} a^{i_2 )} | 0 \rangle & = & | \twobox \rangle, \nn \\
~ & \vdots & ~ \nn \\
a^{( i_1} a^{i_2} \ldots a^{i_n )} | 0 \rangle & = & | \marcnbox \rangle, \enn (plus
those obtained by interchanging $a$-type oscillators with $b$-type oscillators) and  the
state \eq a^{(i} b^{j)} | 0 \rangle = | \twobox \rangle. \en These lowest weight vectors
$| \Omega \rangle$ of the doubleton UIRs of $SO^{*}(8)$ all transform in the symmetric
tensor representations of $SU(4)$.

On the other hand, in $d = 6$ conformal field theories
one would like to work with fields that transform covariantly under the Lorentz group
$SU^*(4)$ with a definite conformal dimension. The conformal group $SO(6,2)$ has a
3-graded structure with respect to its subgroup $SU^{*}(4) \times \mathcal{D}$ as well,
where $\mathcal{D}$ is the Abelian group of scale transformations \cite{mgm, mgst}. We
shall refer to this subgroup as the homogeneous Weyl group in $d =
6$.

When $G \equiv
SO(6,2)$ acts in the standard way on the (conformal compactification of) six dimensional
Minkowski spacetime, the stability group $H$ of the coordinate six-vector $x^{\mu} = 0$
is simply the semi-direct product $( SU^{*}(4) \times \mathcal{D} ) \odot \mathcal{K}_6$,
where $\mathcal{K}_6$ represents the Abelian subgroup generated by the special conformal
generators $K_{\mu}$. The conformal fields in $d = 6$ live on the coset space $G/H$.
These fields are labeled by their transformation properties under the Lorentz group
$SU^*(4)$, their conformal dimension $l$ and certain matrices $\kappa_{\mu}$ that
describe their behavior under special conformal transformations $K_{\mu}$ \cite{mgst}.
This is identical to conformal fields in $d = 4$
\cite{macksalam,gmz2}.

To establish a
dictionary between the compact (Wigner picture) and non-compact (Dirac picture) bases of
positive energy representations of $SO(6,2)$, its generators were expressed in terms of
bosonic oscillators transforming in the left-handed spinor representation of $SO(6,2)$ in
\cite{mgst}, which we summarize below.

Consider the $d = 6$ gamma matrices $\Gamma_{\mu}$
satisfying \eq \{ \Gamma_{\mu} , \Gamma_{\nu} \} = 2 \eta_{\mu\nu}, \en with $\Gamma_{7}
= - \Gamma_{0} \Gamma_{1} \Gamma_{2} \Gamma_{3} \Gamma_{4} \Gamma_{5}$. Then the matrices
\eqn
\Sigma_{\mu\nu} & := & \frac{i}{4} \left[ \Gamma_{\mu} , \Gamma_{\nu} \right], \nn \\
\Sigma_{\mu 6} & := & - \frac{1}{2} \Gamma_{\mu} \Gamma_{7}, \nn \\
\Sigma_{\mu 7} & := & \frac{1}{2} \Gamma_{\mu}, \nn \\
\Sigma_{67} & := & \frac{i}{2} \Gamma_{7}, \enn generate the eight dimensional
left-handed spinor representation of the conformal  algebra $SO(6,2)$\footnote{Our choice
of gamma matrices and our conventions are outlined in Appendix B.}.

Let \eq \Psi(r) :=
\left( \begin{array}{c} a_{i}(r) \\ b^{j}(r) \end{array} \right), \en and therefore \eqn
\bar{\Psi}(r) & \equiv & {\Psi}^{\dagger}(r) \Gamma_{0} \nn \\
& = & \left( \begin{array}{cc} a^{i}(r) & - b_{j}(r) \end{array}
\right). \enn

If we
denote the components of the spinor $\Psi$ with lower indices $\Psi_{A}$ and  the
components of the Dirac conjugate spinor $\bar{\Psi}$ with upper indices $\bar{\Psi}^{B}$
($A,B = 1,2,\ldots,8$), they satisfy \eq \left[ \Psi_{A}(r) , \overline{\Psi}^{B}(s)
\right] = \delta_{A}^{~B} \delta_{rs}. \en

The bilinears of these twistorial operators
involving the $8 \times 8$ matrices $\Sigma_{ab}$ : \eq \bar{\Psi} \Sigma_{ab} \Psi :=
\sum_{r=1}^{P} \bar{\Psi}(r) \Sigma_{ab} \Psi(r), \en satisfy the commutation relations
of $SO(6,2)$, \eq \left[ \bar{\Psi} \Sigma_{ab} \Psi , \bar{\Psi} \Sigma_{cd} \Psi
\right] = \bar{\Psi} \left[ \Sigma_{ab} , \Sigma_{cd} \right] \Psi, \en and yield
infinite dimensional unitary representations of $SO(6,2)$ in the Fock space of the
oscillators $a^{i}(r)$ and $b^{j}(r)$ \cite{mgst}.\footnote{In this paper when we write
the generators of a Lie (super)algebra, we shall assume that the color indices are summed
over, and drop the summation symbol and the color indices.}

The generators of the
conformal algebra in $d = 6$ can be written as \eqn
M_{\mu \nu} & = & \frac{i}{4} \bar{\Psi}^{B} \left[ \Gamma_{\mu} ,
 \Gamma_{\nu} \right]_{B}^{~A} \Psi_{A} , \nn \\
D & = & - \frac{i}{2} \bar{\Psi}^{B} \left( \Gamma_{7} \right)_{B}^{~A} \Psi_{A} , \nn \\
P_{\mu} & = & \frac{1}{2} \bar{\Psi}^{B} \left( \Gamma_{\mu} ( I - \Gamma_{7} )
 \right)_{B}^{~A} \Psi_{A} , \nn \\
K_{\mu} & = & \frac{1}{2} \bar{\Psi}^{B} \left( \Gamma_{\mu} ( I + \Gamma_{7} )
\right)_{B}^{~A} \Psi_{A} . \enn

Triality, viz. the existence of left-handed spinor,
right-handed spinor and vector  representations which are all eight-dimensional, allows
one to write the generators of $SO(6,2)$ as anti-symmetric tensors in the spinor
representation which satisfy the commutation relations \eq \left[ \tilde{M}_{A B} ,
\tilde{M}_{C D} \right] = \frac{1}{2} \left( C_{B C} \tilde{M}_{A D} - C_{A C}
\tilde{M}_{B D} - C_{B D} \tilde{M}_{A C} + C_{A D} \tilde{M}_{B C} \right), \en where
\eq \tilde{M}_{A B} = \frac{1}{2} \left( \bar{\Psi}^{C} C_{C A} \Psi_{B} - \bar{\Psi}^{C}
C_{C B} \Psi_{A} \right), \label{SOM} \en and $C_{A B}$ is the charge conjugation matrix
in six dimensions, which is symmetric.

It should be noted that the hermitian conjugate of
$\tilde{M}_{A B}$ can be expressed as \eq \left( \tilde{M}_{A B} \right)^{\dag} \equiv
\tilde{M}^{A B} = C^{A C} C^{B D} \tilde{M}_{C D}. \en

The positive energy UIRs of
$SO^{*}(8)$ can be identified with conformal fields in $d = 6$, with a definite conformal
dimension, transforming covariantly under the six dimensional Lorentz group and with
trivial special conformal transformations. To establish this connection, we need to find
a mapping from the $SU^{*}(4)$- and $\mathcal{D}$- covariant basis to the compact $U(4)$
basis.

For this, one introduces the operator \eq U := e^{\bar{\Psi} ( \Sigma_{06} + i
\Sigma_{67} ) \Psi}, \en which satisfies the following important relations : \eqn
M_{mn} U & = & U M_{mn} \hspace{2cm} \mbox{for } m,n = 1,2,\ldots,5, \nn \\
i M_{m0} U & = & U ( M_{m6} + L^{-} ), \nn \\
i D U & = & U ( E + L^{-} ), \nn \\
K_{\mu} U & = & U L^{-}, \label{UL} \enn where $L^{-}$ stands for certain linear
combinations of the di-annihilation  operators $A_{ij}$, whose explicit form is different
for the three equations above. Thus $U$ can be considered as the ``intertwiner'' between
the generators $( M_{\mu \nu}, D )$ of the Lorentz group and dilatations and the
generators $(M_{mn},E )$ of the maximal compact subgroup $SU(4) \times U(1)$. The indices
$m$,$n$ above are the $SO(6)$ vector indices.

In four dimensions, the finite dimensional
representations of the Lorentz group $SL(2,C)$ are labeled as $(j_{1} , j_{2})$ of the
Wick rotated compact Lorentz group $SU(2) \times SU(2)$. In analogy with four
dimensions, we can define a compact Wick rotated Lorentz group, $SU_{c}^{*}(4)$ whose
generators are \eq J_{\hat{m} \hat{n}} = \left\{ \begin{array}{ll} M_{\hat{m} \hat{n}} &
\hat{m},\hat{n} = 1,2,\ldots,5, \\ i M_{\hat{m} 0} & \hat{m} = 1,2,\ldots,5; \hat{n} = 6.
\\ \end{array} \right. \en

The common subgroup of $SU_{c}^{*}(4)$ and the compact
subgroup $SU(4)$ is the rotation group $USp(4) (\cong SO(5))$.

Acting with $U$ on a
lowest weight vector $| \Omega \rangle$ corresponds to a (complex) rotation in the
corresponding representation space of $SO^{*}(8)$ : \eq U | \Omega \rangle =
e^{\bar{\Psi} ( \Sigma_{06} + i \Sigma_{67} ) \Psi} | \Omega
\rangle. \en

For any lowest
weight vector $| \Omega \rangle$ in the compact basis, which transforms irreducibly under
the compact subgroup $SU(4)$, the state $U | \Omega \rangle$ transforms in the same
irreducible representation of $SU_{c}^{*}(4)$. Now the states $| \Omega \rangle$ are
created by the action of $SU(4)$ oscillators $a^{i}$ and $b^{j}$ on $| 0 \rangle$.
Correspondingly, one can define $SU_{c}^{*}(4)$ covariant oscillators $A^{\hat{i}}$ and
$B^{\hat{j}}$ that create the states $U | \Omega \rangle$ by acting on $U | 0 \rangle$ :
\eqn
A^{\hat{i}} (r) U | 0 \rangle & \propto & U a^{i} (r) | 0 \rangle, \nn \\
B^{\hat{j}} (r) U | 0 \rangle & \propto & U b^{j} (r) | 0 \rangle, \enn (up to a possible
normalization constant from (\ref{ABCOMM})).  The respective annihilation operators
$A_{\hat{i}}$ and $B_{\hat{j}}$ are chosen such that, \eqn
A_{\hat{i}} (r) U | 0 \rangle & \propto & U a_{i} (r) | 0 \rangle \hspace{0.2cm}
= \hspace{0.2cm} 0, \nn \\
B_{\hat{j}} (r) U | 0 \rangle & \propto & U b_{j} (r) | 0 \rangle \hspace{0.2cm}
 = \hspace{0.2cm} 0.
\enn
We further require them to satisfy the following commutation relations,
\eqn
\left[ A_{\hat{i}} (r) , A^{\hat{j}} (s) \right] & = & \delta_{\hat{i}}^{~\hat{j}} \delta_{rs},
\nn \\
\left[ B_{\hat{i}} (r) , B^{\hat{j}} (s) \right] & = & \delta_{\hat{i}}^{~\hat{j}} \delta_{rs},
 \label{ABCOMM}
\enn while all the other commutators among them vanish. These oscillators  $A_{\hat{i}}$,
$B_{\hat{j}}$, $A^{\hat{i}}$, $B^{\hat{j}}$ transform covariantly with respect to
$SU_{c}^{*}(4)$.

Remarkably, one finds that $L^{-} | \Omega \rangle = 0$ implies that
$K_{\mu} U | \Omega \rangle = 0$ \cite{mgst}. Thus, every unitary lowest weight
representation (ULWR) of $SO^{*}(8)$ can be identified with a unitary representation of
$SO(6,2)$ induced by a finite dimensional irreducible representation of $SU^{*}(4)$
(labeled by $SU_{c}^{*}(4)$ Dynkin labels), with a definite conformal dimension $l$ and
trivially realized $K_{\mu}$.

$(SU^{*}(4) \times \mathcal{D}) \odot \mathcal{K}_{6}$ is
the stability group of the coordinate vector $x_{\mu} = 0$. To generate a state at any
other point in spacetime, we need to act with the translation operator as shown below :
\eq e^{i x^{\mu} P_{\mu}} U | \Omega \rangle = | \Phi_{( d_{1}, d_{2}, d_{3} )}(x)
\rangle, \en where $(d_{1}, d_{2}, d_{3})$ are the Dynkin labels\footnote{Our definition
of Dynkin labels is such that, the fundamental representation corresponds to $(1,0,0)$.}
of the irreducible representations of $SU(4)$ and $SU_{c}^{*}(4)$ under which $| \Omega
\rangle$ and $U | \Omega \rangle$ transform, respectively. Thus every irreducible ULWR of
$SO(6,2)$ corresponds to a conformal field, that transforms covariantly under $SU^{*}(4)$
with a definite conformal dimension $l = -E$.

We recall that the doubleton
representations of $SO^{*}(8)$ correspond to taking a single pair ($P = 1$) of bosonic
oscillators and that they do not have a smooth Poincar\'{e} limit in $d = 7$. Consider
the Poincar\'{e} mass operator \eq M^{2} = P_{\mu} P^{\mu} \en in $d = 6$ Minkowski
spacetime, where the translation generators $P_{\mu}$ have the following realization in
terms of the $SU(4)$ covariant oscillators : \eqn
\begin{array}{ll}
P_{0} & = \frac{1}{2} \left\{ ( a^{1} - b_{3} ) ( a_{1} - b^{3} ) + ( a^{2} + b_{4} )
( a_{2} + b^{4} ) + ( a^{3} + b_{1} ) ( a_{3} + b^{1} ) + ( a^{4} - b_{2} )
( a_{4} - b^{2} ) \right\}, \nn \\
P_{1} & = \frac{1}{2} \left\{ - ( a^{1} - b_{3} ) ( a_{1} - b^{3} ) + ( a^{2} + b_{4} )
( a_{2} + b^{4} ) - ( a^{3} + b_{1} ) ( a_{3} + b^{1} ) + ( a^{4} - b_{2} ) ( a_{4} - b^{2} )
\right\}, \nn \\
P_{2} & = \frac{1}{2} \left\{ ( a^{1} - b_{3} ) ( a_{2} + b^{4} ) + ( a^{2} + b_{4} )
( a_{1} - b^{3} ) - ( a^{3} + b_{1} ) ( a_{4} - b^{2} ) - ( a^{4} - b_{2} ) ( a_{3} + b^{1} )
 \right\}, \nn \\
P_{3} & = \frac{1}{2} \left\{ - ( a^{1} - b_{3} ) ( a_{4} - b^{2} ) - ( a^{2} + b_{4} )
( a_{3} + b^{1} ) - ( a^{3} + b_{1} ) ( a_{2} + b^{4} ) - ( a^{4} - b_{2} ) ( a_{1} - b^{3} )
 \right\}, \nn \\
P_{4} & = \frac{i}{2} \left\{ ( a^{1} - b_{3} ) ( a_{4} - b^{2} ) + ( a^{2} + b_{4} )
( a_{3} + b^{1} ) - ( a^{3} + b_{1} ) ( a_{2} + b^{4} ) - ( a^{4} - b_{2} ) ( a_{1} - b^{3} )
 \right\}, \nn \\
P_{5} & = \frac{i}{2} \left\{ - ( a^{1} - b_{3} ) ( a_{2} + b^{4} ) + ( a^{2} + b_{4} )
 ( a_{1} - b^{3} ) - ( a^{3} + b_{1} ) ( a_{4} - b^{2} ) + ( a^{4} - b_{2} ) ( a_{3} + b^{1} )
 \right\}.
\end{array}
\enn

Substituting in the above expressions for $P_{\mu}$ one finds that  the mass
operator $M^{2}$ vanishes identically for $P = 1$ \cite{mgst}. Thus all the doubleton
irreducible representations of $SO^*(8)$ are massless in $d = 6$. For $P \neq 1$ the mass
operator is non-vanishing and the corresponding ULWRs of $SO^{*}(8)$ define massive
conformal fields in $d = 6$. We should stress that this is in complete parallel to the
situation in $d = 4$, where the doubleton representations of $SO(4,2)$ are all massless
\cite{binegar,gmz2}.

The doubleton irreducible representations of $SO(6,2)$ and the
corresponding conformal fields are listed in Table 1. \vspace{.2cm}
\begin{center}
\begin{tabular}{|c|c|c|}
\hline
lowest weight vector & $SU^*(4)$ field labels & conformal dimension \\
& $| \Phi_{( d_{1}, d_{2}, d_{3} )}(x) \rangle$ & $l$ \\ \hline \hline
$U | 0 \rangle$ & $| \Phi_{(0,0,0)}(x) \rangle$ & $-2$ \\ \hline
$A^{\hat{i}_{1}} U | 0 \rangle$ & $| \Phi_{(1,0,0)}(x) \rangle$ & $- \frac{5}{2}$ \\ \hline
$A^{( \hat{i}_{1}} A^{\hat{i}_{2} )} U | 0 \rangle$ & $| \Phi_{(2,0,0)}(x) \rangle$ & $-3$ \\
\hline
\vdots & \vdots & \vdots \\ \hline
$A^{( \hat{i}_{1}} \ldots A^{\hat{i}_{n} )} U | 0 \rangle$ & $| \Phi_{(n,0,0)}(x)
 \rangle$ & $- \frac{1}{2}(n+4)$ \\ \hline
$A^{( \hat{i}_{1}} B^{\hat{j}_{2} )} U | 0 \rangle$ & $| \Phi_{(2,0,0)}(x)
\rangle$ & $-3$ \\ \hline
\end{tabular}
\end{center}
\begin{center} Table 1. Possible lowest weight vectors of doubleton representations,
 corresponding conformal fields and their conformal dimensions. \end{center}
\vspace{.2cm}

The $SU_{c}^{*}(4)$ covariant oscillators $A_{\hat{i}}$, $B_{\hat{j}}$, $A^{\hat{i}}$,
 $B^{\hat{j}}$ can be expressed in terms of $a_{i}$, $b_{j}$, $a^{i}$, $b^{j}$ as
 follows\footnote{Here we omit the color index.} :
\eqn
A_{1} = \frac{1}{\sqrt{2}} ( a_{1} + b^{3} ) \hspace{2cm} B_{1} =
\frac{1}{\sqrt{2}} ( b_{1} - a^{3} ) \nn \\
A_{2} = \frac{1}{\sqrt{2}} ( a_{2} - b^{4} ) \hspace{2cm} B_{2} =
\frac{1}{\sqrt{2}} ( b_{2} + a^{4} ) \nn \\
A_{3} = \frac{1}{\sqrt{2}} ( a_{3} - b^{1} ) \hspace{2cm} B_{3} =
\frac{1}{\sqrt{2}} ( b_{3} + a^{1} ) \nn \\
A_{4} = \frac{1}{\sqrt{2}} ( a_{4} + b^{2} ) \hspace{2cm} B_{4} =
\frac{1}{\sqrt{2}} ( b_{4} - a^{2} ) \nn \\
A^{1} = \frac{1}{\sqrt{2}} ( a^{1} - b_{3} ) \hspace{2cm} B^{1} =
\frac{1}{\sqrt{2}} ( b^{1} + a_{3} ) \nn \\
A^{2} = \frac{1}{\sqrt{2}} ( a^{2} + b_{4} ) \hspace{2cm} B^{2} =
\frac{1}{\sqrt{2}} ( b^{2} - a_{4} ) \nn \\
A^{3} = \frac{1}{\sqrt{2}} ( a^{3} + b_{1} ) \hspace{2cm} B^{3} =
\frac{1}{\sqrt{2}} ( b^{3} - a_{1} ) \nn \\
A^{4} = \frac{1}{\sqrt{2}} ( a^{4} - b_{2} ) \hspace{2cm} B^{4} =
 \frac{1}{\sqrt{2}} ( b^{4} + a_{2} ).
\enn

Note that for our $SU_{c}^{*}(4)$ covariant oscillators,  $A^{\hat{i}} \neq
(A_{\hat{i}})^{\dag}$ and $B^{\hat{j}} \neq (B_{\hat{j}})^{\dag}$ with respect to the
standard conjugation $\dag$ in the Fock space of $SU(4)$ covariant oscillators $a_{i}$,
$b_{j}$, $a^{i}$, $b^{j}$. In fact, \eqn
\begin{array}{llll}
\left( A_{1} \right)^{\dag} = B_{3} \hspace{1cm} & \left( A_{2} \right)^{\dag} = - B_{4}
 \hspace{1cm} & \left( A_{3} \right)^{\dag} = - B_{1} \hspace{1cm} & \left( A_{4} \right)^{\dag}
  = B_{2} \\
\left( B^{1} \right)^{\dag} = A^{3} \hspace{1cm} & \left( B^{2} \right)^{\dag} = - A^{4}
 \hspace{1cm} & \left( B^{3} \right)^{\dag} = - A^{1} \hspace{1cm} & \left( B^{4} \right)^{\dag}
  = A^{2}.
\end{array}
\enn

The generators of $SO(6,2)$ in the non-compact basis $SU^{*}(4) \times \mathcal{D}$
can then be written in the form \eqn
M_{\mu \nu} & = & \frac{i}{8} \left( \overline{\Upsilon}^{B} \left[ \Gamma_{\mu} ,
 \Gamma_{\nu} \right]_{B}^{~A} \Xi_{A} + \overline{\Xi}^{B} \left[ \Gamma_{\mu} , \Gamma_{\nu}
 \right]_{B}^{~A} \Upsilon_{A} \right), \nn \\
D & = & - \frac{i}{4} \left( \overline{\Upsilon}^{B} ( \Gamma_{7} )_{B}^{~A} \Xi_{A} +
 \overline{\Xi}^{B} ( \Gamma_{7} )_{B}^{~A} \Upsilon_{A} \right), \nn \\
P_{\mu} & = & \frac{1}{2} \overline{\Xi}^{B} ( \Gamma_{\mu} )_{B}^{~A} \Xi_{A}, \nn \\
K_{\mu} & = & \frac{1}{2} \overline{\Upsilon}^{B} ( \Gamma_{\mu} )_{B}^{~A} \Upsilon_{A},
\enn
where
\eqn
\Upsilon_{A} = \frac{1}{\sqrt{2}} \left( I + \Gamma_{7} \right)_{A}^{~B} \Psi_{B} & =
& \frac{1}{\sqrt{2}} \left( \begin{array}{r} a_{1} + b^{3} \\ a_{2} - b^{4} \\ a_{3} - b^{1} \\
a_{4} + b^{2} \\ - a_{3} + b^{1} \\ a_{4} + b^{2} \\ a_{1} + b^{3} \\ - a_{2} + b^{4} \end{array}
\right) = \left( \begin{array}{r} A_{1} \\ A_{2} \\ A_{3} \\ A_{4} \\ - A_{3} \\ A_{4}
 \\ A_{1} \\
 - A_{2} \end{array} \right), \label{upsilon} \\
\Xi_{A} = \frac{1}{\sqrt{2}} \left( I - \Gamma_{7} \right)_{A}^{~B} \Psi_{B} & =
 & \frac{1}{\sqrt{2}} \left( \begin{array}{r} a_{1} - b^{3} \\ a_{2} + b^{4} \\ a_{3} + b^{1} \\
  a_{4} - b^{2} \\ a_{3} + b^{1} \\ - a_{4} + b^{2} \\ - a_{1} + b^{3} \\ a_{2} + b^{4} \end{array} \right) = \left( \begin{array}{r} - B^{3} \\ B^{4} \\ B^{1} \\ - B^{2} \\ B^{1} \\ B^{2} \\ B^{3} \\ B^{4} \end{array} \right), \label{xi}
\enn
and hence
\eqn
\overline{\Upsilon}^{A} & = & \frac{1}{\sqrt{2}} \overline{\Psi}^{B} \left( I - \Gamma_{7} \right)
 _{B}^{~A} \nn \\
& = & \frac{1}{\sqrt{2}} \left( \begin{array}{cccccccc} a^{1} + b_{3} & a^{2} - b_{4} & a^{3}
 - b_{1} & a^{4} + b_{2} & a^{3} - b_{1} & - a^{4} - b_{2} & - a^{1} - b_{3} & a^{2} - b_{4}
  \end{array} \right) \nn \\
& = & \left( \begin{array}{cccccccc} B_{3} & - B_{4} & - B_{1} & B_{2} & - B_{1} & - B_{2} &
 - B_{3} & - B_{4} \end{array} \right), \\
\overline{\Xi}^{A} & = & \frac{1}{\sqrt{2}} \overline{\Psi}^{B} \left( I + \Gamma_{7}
\right)_{B}^{~A} \nn \\
& = & \frac{1}{\sqrt{2}} \left( \begin{array}{cccccccc} a^{1} - b_{3} & a^{2} + b_{4} & a^{3}
 + b_{1} & a^{4} - b_{2} & - a^{3} - b_{1} & a^{4} - b_{2} & a^{1} - b_{3} & - a^{2} - b_{4}
  \end{array} \right) \nn \\
& = & \left( \begin{array}{cccccccc} A^{1} & A^{2} & A^{3} & A^{4} & - A^{3} & A^{4} & A^{1}
 & - A^{2} \end{array} \right).
\enn

These $SU_{c}^{*}(4)$ spinors satisfy the commutation relations,
\eqn
\left[ \Upsilon_{A}(r) , \overline{\Xi}^{B}(s) \right] & = & ( I + \Gamma_{7} )_{A}^{~B}
\delta_{rs}, \nn \\
\left[ \Xi_{A}(r) , \overline{\Upsilon}^{B}(s) \right] & = & ( I - \Gamma_{7} )_{A}^{~B}
 \delta_{rs},
\enn
while all the other commutators vanish.

In terms of these $SU_{c}^{*}(4)$ covariant oscillators, $P_{\mu}$ and $K_{\mu}$ are purely
 di-creation and di-annihilation operators, respectively, and as a result, the proof that
 the conformal fields associated with the doubleton irreducible representations are all
  massless in $d = 6$ \cite{mgst}, is greatly simplified when these covariant oscillators
   are used. The realization of $P_{\mu}$ in terms of these oscillators is given below.
\eqn
P_{0} & = & - A^{1} B^{3} + A^{2} B^{4} + A^{3} B^{1} - A^{4} B^{2}, \nn \\
P_{1} & = & A^{1} B^{3} + A^{2} B^{4} - A^{3} B^{1} - A^{4} B^{2}, \nn \\
P_{2} & = & A^{1} B^{4} - A^{2} B^{3} + A^{3} B^{2} - A^{4} B^{1}, \nn \\
P_{3} & = & A^{1} B^{2} - A^{2} B^{1} - A^{3} B^{4} + A^{4} B^{3}, \nn \\
P_{4} & = & i \left( - A^{1} B^{2} + A^{2} B^{1} - A^{3} B^{4} + A^{4} B^{3} \right), \nn \\
P_{5} & = & i \left( - A^{1} B^{4} - A^{2} B^{3} + A^{3} B^{2} + A^{4} B^{1} \right).
\enn

Similarly, one finds the following expressions for special conformal transformations
 $K_{\mu}$ and dilatation generator $D$ in terms of these $SU_{c}^{*}(4)$ covariant oscillators :
\eqn
K_{0} & = & A_{1} B_{3} - A_{2} B_{4} - A_{3} B_{1} + A_{4} B_{2}, \nn \\
K_{1} & = & A_{1} B_{3} + A_{2} B_{4} - A_{3} B_{1} - A_{4} B_{2}, \nn \\
K_{2} & = & A_{1} B_{4} - A_{2} B_{3} + A_{3} B_{2} - A_{4} B_{1}, \nn \\
K_{3} & = & A_{1} B_{2} - A_{2} B_{1} - A_{3} B_{4} + A_{4} B_{3}, \nn \\
K_{4} & = & i \left( A_{1} B_{2} - A_{2} B_{1} + A_{3} B_{4} - A_{4} B_{3} \right), \nn \\
K_{5} & = & i \left( A_{1} B_{4} + A_{2} B_{3} - A_{3} B_{2} - A_{4} B_{1} \right), \enn
\eqn D = - \frac{i}{2} \left( A^{1} A_{1} + A^{2} A_{2} + A^{3} A_{3} + A^{4} A_{4} +
B_{1} B^{1} + B_{2} B^{2} + B_{3} B^{3} + B_{4} B^{4} \right). \enn

The massless
representations of $SO^{*}(8)$, considered as  the seven dimensional $AdS$ group, are
obtained by taking $P = 2$ pairs of oscillators. However, as representations of the
conformal group in $d = 6$, they are massive.

In Table 2, we give these irreducible
representations and their corresponding conformal fields. \vspace{.2cm}
\begin{center}
\begin{tabular}{|c|c|c|}
\hline
lowest weight vector & $SU^*(4)$ field labels & conformal dimension \\
& $| \Phi_{( d_{1}, d_{2}, d_{3} )}(x) \rangle$ & $l$ \\ \hline \hline
$U | 0 \rangle$ & $| \Phi_{(0,0,0)}(x) \rangle$ & -4 \\ \hline
$A^{\hat{i}_{1}}(r) U|0 \rangle$ & $| \Phi_{(1,0,0)}(x) \rangle$ & $- \frac{9}{2}$ \\ \hline
$A^{( \hat{i}_{1}}(r) A^{\hat{i}_{2} )}(r) U | 0 \rangle$ & $| \Phi_{(2,0,0)}(x)
\rangle$ & $-5$ \\
$A^{( \hat{i}_{1}}(r) A^{\hat{i}_{2} )}(s) U | 0 \rangle$ & & \\ \hline
$A^{[ \hat{i}_{1}}(r) A^{\hat{i}_{2} ]}(s) U | 0 \rangle$ & $| \Phi_{(0,1,0)}(x)
\rangle$ & $-5$ \\ \hline
$A^{( \hat{i}_{1}}(r) B^{\hat{j}_{1} )}(r) U | 0 \rangle$ & $| \Phi_{(2,0,0)}(x)
 \rangle$ & $-5$ \\
$A^{( \hat{i}_{1}}(r) B^{\hat{j}_{1} )}(s) U | 0 \rangle$ & & \\ \hline
$A^{[ \hat{i}_{1}}(r) B^{\hat{j}_{1} ]}(s) U | 0 \rangle$ & $| \Phi_{(0,1,0)}(x)
\rangle$ & $-5$ \\ \hline
\vdots & \vdots & \vdots \\
\vdots & \vdots & \vdots \\ \hline
$A^{( \hat{i}_{1}}(r) \ldots A^{\hat{i}_{n} )}(r) U |0 \rangle$ & $| \Phi_{(n,0,0)}(x) \rangle$
& $- \frac{1}{2} (n+8)$ \\
$A^{( \hat{i}_{1}}(r) \ldots A^{\hat{i}_{m}}(r) A^{\hat{i}_{m+1}}(s)
 \ldots A^{\hat{i}_{n} )}(s) U |0 \rangle$ & & \\ \hline
$A^{( \hat{i}_{1}}(r) \ldots A^{\hat{i}_{m}}(r) B^{\hat{j}_{m+1}}(s)
 \ldots B^{\hat{j}_{n} )}(s) U |0 \rangle$ & $| \Phi_{(n,0,0)}(x)
 \rangle$ & $- \frac{1}{2} (n+8)$ \\ \hline
$A^{[ \hat{i}_{1}}(r) A^{\hat{j}_{1} ]}(s) \ldots A^{[ \hat{i}_{m}}(r) A^{\hat{j}_{m} ]}(s)
 A^{\hat{i}_{m+1}}(r) \ldots A^{\hat{i}_{m+n}}(r) U|0 \rangle$ & $| \Phi_{(n,m,0)}(x)
 \rangle$ & $- \frac{1}{2} (2m+n+8)$ \\ \hline
$A^{[ \hat{i}_{1}}(r) B^{\hat{j}_{1} ]}(s) \ldots A^{[ \hat{i}_{m}}(r) B^{\hat{j}_{m} ]}(s)
 A^{\hat{i}_{m+1}}(r) \ldots A^{\hat{i}_{m+n}}(r) U|0 \rangle$ & $| \Phi_{(n,m,0)}(x)
 \rangle$ & $- \frac{1}{2} (2m+n+8)$ \\ \hline
\end{tabular}
\end{center}
\begin{center} Table 2. Possible lowest weight vectors of $SO^{*}(8)$ for $P = 2$,
corresponding massive conformal fields in $d = 6$ and their conformal dimensions. Above,
 the color indices $r,s = 1,2$ and $r \neq s$. \end{center}
\vspace{.2cm}

Similarly, for $P > 2$ one obtains representations of $SO^{*}(8)$, which are massive, both
 as $d = 6$ conformal fields and as $AdS_{7}$ fields.

For $P = 3$, the possible lowest weight vectors $| \Omega \rangle$ ($U | \Omega \rangle$)
 can be in any representation of $SU(4)$ ($SU_{c}^{*}(4)$). If $| \Omega \rangle$ ($U |
 \Omega \rangle$) transforms in the representation $(d_{1}, d_{2}, d_{3})_{D}$, then the
 corresponding conformal field $| \Phi_{(d_{1},d_{2},d_{3})} (x) \rangle$ has the conformal
  dimension
\eq
l = - \frac{1}{2} ( d_{1} + 2 d_{2} + 3 d_{3} + 12 ).
\en

For $P \geq 4$, there can be $SU(4)$ singlet lowest weight vectors in addition to the vacuum
 $| 0 \rangle$ (\emph{e.g.} $| \Omega \rangle = \epsilon_{ijkl} a^{i}(r_{1}) a^{j}(r_{2})
 a^{k}(r_{3}) a^{l}(r_{4}) | 0 \rangle$). In this case, it is convenient to use the Young
 Tableaux of the lowest weight vectors with respect to $U(4) = SU(4) \times U(1)$. If we denote
 the Young Tableaux of $U(4)$ by $( m_{1}, m_{2}, m_{3}, m_{4} )_{YT}$, then the corresponding
  conformal fields $| \Phi_{(d_{1}, d_{2}, d_{3})} (x) \rangle$, transform in the
   representation $( d_{1}=m_{1}-m_{2}, d_{2}=m_{2}-m_{3}, d_{3}=m_{3}-m_{4} )_{D}$ of $SU(4)$
    and has the conformal dimension
\eq
l = - \frac{1}{2} ( m_{1} + m_{2} + m_{3} + m_{4} + 4 P ).
\en

Since the bosonic oscillators, in terms of which we realized the generators, transform
in the spinor representation of $SO^{*}(8)$, the oscillator construction can be given a
dynamical realization in terms of twistors as was done for $SU(2,2)$ \cite{cgkrz,ckr}.

\section{Compact versus non-compact bases of the supergroup \OSN}
\setcounter{equation}{0}

The supergroup \OS ~with the even subgroup $SO^{*}(8) \times
USp(4)$ is the symmetry group of M-theory on $AdS_7 \times S^4$. One can interpret \OS
~either as the $\mathcal{N} = 4$ extended $AdS$ superalgebra in $d=7$ or as the $(2,0)$
extended conformal superalgebra in $d = 6$ \cite{gnw}. The finite dimensional
representations of $SO(6,2)$ ($\cong SO^{*}(8)$) possess the triality property and the
anti-symmetric tensor of any one of left-handed spinor, right-handed spinor and vector
representations with itself transforms like the adjoint representation of $SO(6,2)$.
Therefore, there exists three different forms of the \OSN
~superalgebra.

The relevant
form of \OS ~for M-theory on $AdS_7 \times S^4$ is the one for which the supersymmetry
generators, ${\bf \mathcal{R}}_{A I}$ transform as the left-handed spinor representation
of $SO(6,2)$, which decomposes as $(4 + \bar{4})$ with respect to the compact subgroup
$SU(4)$ as well as $SU_{c}^{*}(4)$.

The supersymmetry generators ${\bf \mathcal{R}}_{A
I}$ of \OSN ~satisfy the following anti-commutation relation \cite{ckp,mgst} : \eq
\left\{ {\bf \mathcal{R}}_{A I} , {\bf \mathcal{R}}_{B J} \right\} = - \frac{1}{2} \left(
\tilde{M}_{A B} \Omega_{IJ} + C_{A B} U_{IJ} \right), \en where $A,B = 1,2,\ldots,8$ and
$I,J = 1,2,\ldots,2N$. $\tilde{M}_{A B}$ are the $SO(6,2)$ generators given in
(\ref{SOM}). $U_{IJ} = U_{JI}$ are the $USp(2N)$ generators and $\Omega_{IJ} = -
\Omega_{JI}$ is the symplectic invariant tensor \cite{ckp,mgst}. The $USp(2N)$ generators
satisfy \eq \left[ U_{IJ} , U_{KL} \right] = \Omega_{I(K} U_{L)J} + \Omega_{J(K} U_{L)I}.
\en

 One can define fermionic annihilation ($\alpha_{\kappa}$, $\beta_{\lambda}$) and
creation ($\alpha^{\kappa}$, $\beta^{\lambda}$) operators transforming in the fundamental
representation of $U(N)$ and its conjugate, similar to their bosonic counterparts ($a$,
$b$ or $A$, $B$), such that they satisfy the anti-commutation relations \eqn
\left\{ \alpha_{\kappa}(r) , \alpha^{\lambda}(s) \right\} & = & \delta_{\kappa}^{~\lambda}
\delta_{rs}, \nn \\
\left\{ \beta_{\kappa}(r) , \beta^{\lambda}(s) \right\} & = & \delta_{\kappa}^{~\lambda}
 \delta_{rs},
\enn
where $\kappa,\lambda = 1,2,\ldots,N$ and $r,s = 1,2,\ldots,P$, while all the other
anti-commutators vanish. Then the $USp(2N)$ generators $U_{IJ}$ can be realized as
\eq
U_{IJ} = \frac{1}{2} \left( \Lambda^{K} \Omega_{KI} \Lambda_{J} + \Lambda^{K} \Omega_{KJ}
\Lambda_{I} \right),
\en
where
\eqn
\Lambda_{I} (r) & = & \left( \begin{array}{c} \alpha_{\kappa} (r) \\ \beta^{\lambda} (r)
 \end{array} \right), \nn \\
\Lambda^{J} (r) & \equiv & \left( \Lambda_{J} (r) \right)^{\dag} \nn \\
& = & \left( \begin{array}{cc} \alpha^{\kappa} (r) & \beta_{\lambda} (r) \end{array}
\right). \enn

Thus, the supersymmetry generators ${\bf \mathcal{R}}_{A I}$ of \OSN ~have
a realization  in terms of spinors $\Psi$ and $\Lambda$ as, \eq {\bf \mathcal{R}}_{A I} =
\frac{1}{2} \left( \bar{\Psi}^{B} C _{B A} \Lambda_{I} - \Lambda^{J} \Omega_{JI} \Psi_{A}
\right). \en

\subsection{The 3-grading of the superalgebra \OSN}

\OSN ~has a 3-grading with respect to its maximal compact subsuperalgebra $U(4|N)$ as follows :
\eq
OSp(8^*|2N) = A_{MN} \oplus M^{M}_{~N} \oplus A^{MN},
\en
where
\eqn
A_{MN} & = & {\bf \xi}_{M} \cdot {\bf \eta}_{N} - {\bf \eta}_{M} \cdot {\bf \xi}_{N}
 \hspace{0.2cm} = \hspace{0.2cm} A_{ij} \oplus A_{\kappa \lambda} \oplus \mathcal{R}_{i \kappa},
  \\
A^{MN} & = & \left( A_{MN} \right)^{\dag} \hspace{0.2cm} = \hspace{0.2cm} {\bf \eta}^{N}
 \cdot {\bf \xi}^{M} - {\bf \xi}^{N} \cdot {\bf \eta}^{M} \hspace{0.2cm} = \hspace{0.2cm}
  A^{ij} \oplus A^{\kappa \lambda} \oplus \mathcal{R}^{i\kappa}, \\
M^{M}_{~N} & = & {\bf \xi}^{M} \cdot {\bf \xi}_{N} + (-1)^{(degM)(degN)} {\bf \eta}_{N}
 \cdot {\bf \eta}^{M} \hspace{0.2cm} = \hspace{0.2cm} M^{i}_{~j} \oplus M^{\kappa}_{~\lambda}
  \oplus \mathcal{R}^{i}_{~\kappa} \oplus \mathcal{R}_{i}^{~\kappa},
\enn where $degM$ = 0 ($degM$ = 1) if M is a bosonic (fermionic)
index.

The
superoscillators $\xi_{M}$, $\eta_{N}$, $\xi^{M}$, $\eta^{N}$, which transform
covariantly and contravariantly, respectively, under the $U(4|N)$ subsupergroup of \OSN
~are defined as \eqn
\xi_{M}(r) = \left( \begin{array}{c} a_{i}(r) \\ \alpha_{\kappa}(r) \end{array} \right) &
 \hspace{1cm} & \xi^{M}(r) = \left( \begin{array}{c} a^{i}(r) \\ \alpha^{\kappa}(r)
  \end{array} \right), \nn \\
\eta_{N}(s) = \left( \begin{array}{c} b_{j}(s) \\ \beta_{\lambda}(s)
\end{array} \right) & \hspace{1cm} & \eta^{N}(s) = \left( \begin{array}{c} b^{j}(s) \\
 \beta^{\lambda}(s) \end{array} \right),
\enn with $i,j = 1,2,3,4$; $\kappa,\lambda = 1,\ldots,N$ and $r,s = 1,2,\ldots,P$.  They
satisfy the supercanonical commutation relations \eqn
\left\{ \xi_{M}(r) , \xi^{N}(s) \right] & = & \delta_{M}^{~N} \delta_{rs}, \nn \\
\left\{ \eta_{M}(r) , \eta^{N}(s) \right] & = & \delta_{M}^{~N} \delta_{rs}, \enn while
all the other commutators/anti-commutators vanish.

The operators $A_{\kappa \lambda}$,
$A^{\kappa \lambda}$, and $M^{\kappa}_{~\lambda}$  generate the internal symmetry group
$USp(2N)$.

The odd elements of \OSN ~are of the form $\mathcal{R}_{i \kappa}$,
$\mathcal{R}^{i}_{~\kappa}$, $\mathcal{R}_{i}^{~\kappa}$ and $\mathcal{R}^{i \kappa}$, of
which $\mathcal{R}_{i \kappa}$ and $\mathcal{R}^{i \kappa}$ involve only di-annihilation
and di-creation operators, respectively. The ULWRs of \OSN ~are constructed starting from
a lowest weight vector $| \Omega \rangle$, which is annihilated by $A_{MN}$ : \eq A_{MN}|
\Omega \rangle = 0, \label{lwv} \en and transforms irreducibly under $U(4|N)$. Acting on
$| \Omega \rangle$ with $A^{MN}$ repeatedly generates an infinite dimensional basis of
ULWR of \OSN. The irreducibility of the resulting ULWR follows from the irreducibility of
$| \Omega \rangle$ under $U(4|N)$. These lowest weight representations constructed in the
compact basis are manifestly unitary \cite{gnw,mgrs,mgst}.

\subsection{The 5-grading of the superalgebra \OSN}

Recall that $SO(6,2)$, as the $d = 6$ conformal group, has the 3-grading $K_{\mu} \oplus
( M_{\mu \nu} + D ) \oplus P_{\mu}$ with respect to Lorentz group times dilatations.  The
transition from the compact basis of \OSN ~to the non-compact basis requires that we work
with its 5-graded structure. The Poincar\'{e} supersymmetries $Q_{A I}$ ($A =
1,2,\ldots,8$ and $I = 1,2,\ldots,2N$) close into the momentum generators $P_{\mu}$ under
the anti-commutation. Similarly, the special conformal supersymmetries $S_{A I}$ close
into $K_{\mu}$. Hence the superalgebra \OSN ~has a 5-graded decomposition with respect to
the subalgebra $SU^{*}(4) \times \mathcal{D} \times USp(2N)$ : \eqn
\begin{array}{ccccccccccc}
OSp(8^{*}|2N) & = & K_{\mu} & \oplus & S_{A I} & \oplus & [ M_{\mu\nu} + D + U_{IJ} ]
& \oplus & Q_{A I} & \oplus & P_{\mu} \hspace{0.2cm} . \\
& & ( g^{-1} ) & & ( g^{-\frac{1}{2}} ) & & ( g^{0} ) & & ( g^{+\frac{1}{2}} ) & & ( g^{+1} )
\end{array}
\enn

These $Q_{A I}$ and $S_{A I}$ are right-handed (negative chiral) and left-handed
(positive chiral) spinor generators with respect to $SU^{*}(4)$, respectively : \eqn
(\Gamma_{7})_{A}^{~B} Q_{B I} & = & - Q_{A I}, \nn \\
(\Gamma_{7})_{A}^{~B} S_{B I} & = & S_{A I}. \enn

They can be realized in the following
way in terms of the spinors $\Psi_{A}$ ($\bar{\Psi}^{B}$)  and $\Lambda_{I}$
($\Lambda^{J}$) introduced before : \eqn
Q_{A I} & = & \frac{1}{2} \left( I - \Gamma_{7} \right)_{A}^{~B} {\bf \mathcal{R}}_{B I}
\hspace{0.2cm} = \hspace{0.2cm} \frac{1}{4} \left( I - \Gamma_{7} \right)_{A}^{~B}
\left( \bar{\Psi}^{C} C _{C B} \Lambda_{I} - \Lambda^{J} \Omega_{JI} \Psi_{B} \right), \nn \\
S_{A I} & = & \frac{1}{2} \left( I + \Gamma_{7} \right)_{A}^{~B} {\bf \mathcal{R}}_{B I}
 \hspace{0.2cm} = \hspace{0.2cm} \frac{1}{4} \left( I + \Gamma_{7} \right)_{A}^{~B}
 \left( \bar{\Psi}^{C} C _{C B} \Lambda_{I} - \Lambda^{J} \Omega_{JI} \Psi_{B} \right),
\label{qsinpsi}
\enn
or, in the $SU_{c}^{*}(4)$ covariant basis,
\eqn
Q_{A I} & = & \frac{1}{2\sqrt{2}} \left( \overline{\Xi}^{B} C_{B A} \Lambda_{I} -
\Xi_{A} \Lambda^{J} \Omega_{JI} \right), \nn \\
S_{A I} & = & \frac{1}{2\sqrt{2}} \left( \overline{\Upsilon}^{B} C_{B A} \Lambda_{I} -
\Upsilon_{A} \Lambda^{J} \Omega_{JI} \right).
\enn

It is important to note that in this realization, the only bosonic oscillators in
 $Q_{A I}$ are $A^{\hat{i}}$ and $B^{\hat{j}}$, while the only bosonic oscillators in
  $S_{A I}$ are $A_{\hat{i}}$ and $B_{\hat{j}}$. This is consistent with the 5-graded
   structure of \OSN.

The commutation/anti-commutation relations of $Q_{A I}$ and $S_{A I}$ among themselves
and with conformal generators in $d = 6$ are \cite{bsv} :
\eqn
\left[ Q_{A I} , M_{\mu \nu} \right] & = & \frac{i}{4} \left[ \Gamma_{\mu} ,
 \Gamma_{\nu} \right]_{A}^{~B} Q_{B I}, \nn \\
\left[ S_{A I} , M_{\mu \nu} \right] & = & \frac{i}{4} \left[ \Gamma_{\mu} ,
\Gamma_{\nu} \right]_{A}^{~B} S_{B I}, \nn \\
\left[ Q_{A I} , D \right] & = & \frac{i}{2} Q_{A I}, \nn \\
\left[ S_{A I} , D \right] & = & - \frac{i}{2} S_{A I}, \nn \\
\left[ Q_{A I} , P_{\mu} \right] & = & 0, \nn \\
\left[ S_{A I} , P_{\mu} \right] & = & \left( \Gamma_{\mu} \right)_{A}^{~B} Q_{B I}, \nn \\
\left[ Q_{A I} , K_{\mu} \right] & = & \left( \Gamma_{\mu} \right)_{A}^{~B} S_{B I}, \nn \\
\left[ S_{A I} , K_{\mu} \right] & = & 0, \nn \\
\left\{ Q_{A I} , Q_{B J} \right\} & = & \frac{1}{16} \left( ( I - \Gamma_{7} )
\Gamma^{\mu} \right)_{A B} P_{\mu} \Omega_{I J}, \nn \\
\left\{ S_{A I} , S_{B J} \right\} & = & \frac{1}{16} \left( ( I + \Gamma_{7} )
 \Gamma^{\mu} \right)_{A B} K_{\mu} \Omega_{I J}, \nn \\
\left\{ Q_{A I} , S_{B J} \right\} & = & \frac{i}{16} \left\{ \frac{1}{4}
\left( ( I - \Gamma_{7} ) \left[ \Gamma^{\mu} , \Gamma^{\nu} \right] \right)_{A B} M_{\mu \nu}
 - \left( I - \Gamma_{7} \right)_{A B} D \right\} \Omega_{I J} \nn \\
& & - \frac{1}{4} \left( I - \Gamma_{7} \right)_{A B} U_{I J}. \enn

 We note that any
state that is annihilated by $S_{A I}$ is also annihilated by $K_{\mu}$,  but the
converse is not necessarily true. We also recall that any lowest weight vector $U |
\Omega \rangle$ is annihilated by $K_{\mu}$. On the other hand, the components of $S_{A
I}$ are either di-annihilation operators (for $I = 1,\ldots,N$) or involve bilinears of a
bosonic annihilation and a fermionic creation operator (for $I = N+1,\ldots,2N$).
Therefore, for those components of $S_{A I}$ that are di-annihilation operators, we have
\eq S_{A I} U | \Omega \rangle = 0 \hspace{1cm} \mbox{for $I =
1,\ldots,N$.} \hspace{0.5cm} \mbox{(cf. equation (\ref{lwv}))} \en

Note that, $U | \Omega \rangle$ is not an irreducible representation of $USp(2N)$. To obtain
them, one needs to act on $U | \Omega \rangle$ with the operators $\alpha^{( \kappa}
\beta^{\lambda )}$ repeatedly. The resulting irreducible representations of $USp(2N)$ are
labeled by the $U(N)$ labels of $U | \Omega \rangle$. If we denote the irreducible
representation of $USp(2N)$ defined by the lowest weight vector $| \Omega \rangle$ as $|
\Pi \rangle$, then it satisfies, \eq K_{\mu} U | \Pi \rangle = 0. \en Therefore, $e^{i
x^{\mu} P_{\mu}} U | \Pi \rangle$ form the coherent state basis of a ULWR of $SO(6,2)$
transforming in an irreducible representation of $USp(2N)$. We should note that $U | \Pi
\rangle$ in general is not annihilated by all $S_{A I}$.

The analog of the lowest weight
vector $| \Omega \rangle$ in the compact basis is the ``chiral primary state'' $| \zeta
\rangle$ in the non-compact basis, that is annihilated by all the $S_{A I}$ ($I =
1,2,\ldots,2N$) and that transforms irreducibly under $SU^{*}(4) \times \mathcal{D}
\times USp(2N)$. By acting on $| \zeta \rangle$ with the translation operator $e^{i
x^{\mu} P_{\mu}}$, one generates a coherent state corresponding to the chiral primary
field and furthermore, the action of Poincar\'{e} supersymmetry generators $Q_{A I}$ on
$e^{i x^{\mu} P_{\mu}} | \zeta \rangle$ generates the supermultiplet of conformal fields.
The state $| \zeta \rangle$ uniquely defines the supermultiplet and as will become
evident later, there exists such a chiral primary state for every ULWR.

\section{Supercoherent states of \OSN ~and superfields}
\setcounter{equation}{0}

Since $\Gamma_{7}$ is not diagonal in our work, to find the
components of $\Psi_{A}$ that  transform as left-handed and right-handed spinors of
$SU^{*}(4)$, we need to act with the projection operators $\frac{1}{\sqrt{2}} ( I \pm
\Gamma_{7} )$ on $\Psi$. One then finds that the first four components of $\Upsilon_{A}$
(equation (\ref{upsilon})) and the last four components of $\Xi_{A}$ (equation
(\ref{xi})) transform as left-handed and right-handed $SU^{*}(4)$ spinors, respectively.

More explicitly, the $SU^{*}(4)$ left-handed spinor indices $\alpha,\beta = 1,2,3,4$ and
right-handed spinor indices $\dot{\alpha},\dot{\beta} = 1,2,3,4$ correspond to the
$SO^{*}(8)$ left-handed spinor indices $A,B = 1,2,3,4$ and $A,B = 5,6,7,8$, respectively.

Let \eqn
u_{\dot{\alpha}} = i \left( \begin{array}{c} A^{1} \\ A^{2} \\ A^{3} \\ A^{4} \end{array} \right), & \hspace{1cm} & s^{\beta} = i \left( \begin{array}{c} B^{1} \\ B^{2} \\ B^{3} \\ B^{4} \end{array} \right), \nn \\
t_{\alpha} = - i \left( \begin{array}{c} A_{1} \\ A_{2} \\ A_{3} \\ A_{4} \end{array} \right), & \hspace{1cm} & v^{\dot{\beta}} = - i \left( \begin{array}{c} B_{1} \\ B_{2} \\ B_{3} \\ B_{4} \end{array} \right). \label{stuvspinors}
\enn

They satisfy
\eqn
\left[ t_{\alpha} , u_{\dot{\alpha}} \right] & = & \delta_{\alpha \dot{\alpha}}, \nn \\
\left[ v^{\dot{\beta}} , s^{\beta} \right] & = & \delta^{\dot{\beta} \beta},
\enn
while all the other commutators are zero.

Then we find that
\eqn
P_{\mu} & = & s^{\alpha} \left( \Sigma_{\mu} \right)_{\alpha}^{~\dot{\beta}}
u_{\dot{\beta}}, \nn \\
K_{\mu} & = & v^{\dot{\alpha}} \left( \overline{\Sigma}_{\mu} \right)_{\dot{\alpha}}^{~\beta}
 t_{\beta}, \label{pksigma}
\enn
where $\overline{\Sigma}_{\mu} = ( \Sigma_{0}, - \Sigma_{1}, - \Sigma_{2}, - \Sigma_{3},
 - \Sigma_{4}, - \Sigma_{5})$.\footnote{The explicit form of the $\Sigma_{\mu}$ matrices
 is given in Appendix C.} These $\Sigma$-matrices in $d = 6$ are the analogs of Pauli
  matrices $\sigma_{\mu}$ in $d = 4$.

Note that under the standard hermitian conjugation over the Fock space of $SU(4)$
covariant oscillators, we have \eqn
\left( t_{\alpha} \right)^{\dag} & = & t_{\dot{\alpha}} \nn \\
\left( s^{\beta} \right)^{\dag} & = & s^{\dot{\beta}}.
\enn

Further, we define
\eqn
\overline{t}_{\alpha} = c_{\alpha}^{~\dot{\beta}} t_{\dot{\beta}}, & \hspace{1cm} &
 \overline{s}^{\beta} = c^{\beta}_{~\dot{\alpha}} s^{\dot{\alpha}}, \nn \\
\overline{t}_{\dot{\alpha}} = t_{\beta} c^{\beta}_{~\dot{\alpha}}, & \hspace{1cm} &
 \overline{s}^{\dot{\beta}} = s^{\alpha} c_{\alpha}^{~\dot{\beta}},
\enn
where the unitary $c$-matrix \cite{koller,tobee} is chosen as :
\eq
c_{\alpha}^{~\dot{\beta}} = \left( \begin{array}{cccc} 0 & 0 & 1 & 0 \\ 0 & 0 & 0 & -1
\\ -1 & 0 & 0 & 0 \\ 0 & 1 & 0 & 0 \end{array} \right) \hspace{2.5cm} c^{\beta}_{~\dot{\alpha}}
 = \left( \begin{array}{cccc} 0 & 0 & -1 & 0 \\ 0 & 0 & 0 & 1 \\ 1 & 0 & 0 & 0 \\
  0 & -1 & 0 & 0 \end{array} \right),
\en
and satisfy
\eqn
c_{\alpha}^{~\dot{\beta}} c_{\dot{\beta}}^{~\gamma} = - \delta_{\alpha}^{~\gamma}, &
 \hspace{1cm} & c_{\dot{\alpha}}^{~\beta} c_{\beta}^{~\dot{\gamma}} =
  - \delta_{\dot{\alpha}}^{~\dot{\gamma}}, \nn \\
\left( c_{\alpha}^{~\dot{\beta}} \right)^{*} = c_{\dot{\alpha}}^{~\beta} = - \left( c^{-1}
\right)_{\dot{\alpha}}^{~\beta}, & \hspace{1cm} & \left( c_{\dot{\alpha}}^{~\beta} \right)^{T}
= c^{\beta}_{~\dot{\alpha}} = \left( c^{-1} \right)_{\dot{\alpha}}^{~\beta}.
\enn

Using the conventions above, it follows that
\eqn
\begin{array}{lll}
\overline{t}_{\alpha} = v^{\dot{\alpha}}, & \hspace{1cm} &
\overline{s}^{\beta} = - u_{\dot{\beta}}, \\
\overline{t}_{\dot{\alpha}} = \left( \overline{t}_{\alpha} \right)^{\dag} =
 \left( v^{\dot{\alpha}} \right)^{\dag} = v^{\alpha}, & \hspace{1cm} &
  \overline{s}^{\dot{\beta}} = \left( \overline{s}^{\beta} \right)^{\dag} = -
  \left( u_{\dot{\beta}} \right)^{\dag} = - u_{\beta}.
\end{array}
\enn

Now, one finds

\eqn
\left( ( I - \Gamma_{7} ) \Gamma^{\mu} \right)^{\alpha \beta} P_{\mu} & = &
4 s^{[ \alpha} \overline{s}^{\beta ]}, \nn \\
\left( ( I + \Gamma_{7} ) \Gamma^{\mu} \right)_{\alpha \beta} K_{\mu} & = &
4 t_{[ \alpha} \overline{t}_{\beta ]}.
\enn

We should note that the $\Sigma$-matrices satisfy the identities

\eqn
\left( \Sigma_{\mu} \right)_{\alpha}^{~\beta} & = & \left( \Sigma_{\mu}
\right)_{\alpha}^{~\dot{\gamma}} c^{\beta}_{~\dot{\gamma}} \hspace{0.2cm} =
 \hspace{0.2cm} - \left( \Sigma_{\mu} \right)_{\alpha}^{~\dot{\gamma}}
  c_{~\dot{\gamma}}^{\beta}, \nn \\
\left( \Sigma_{\mu} \right)_{\alpha}^{~\beta} & = & c_{\alpha}^{~\dot{\gamma}}
 \left( \Sigma_{\mu} \right)_{\dot{\gamma}}^{~\beta}, \nn \\
\left( \Sigma_{\mu} \right)_{\dot{\alpha}}^{~\dot{\beta}} & = & \left( \Sigma_{\mu}
 \right)_{\gamma}^{~\dot{\beta}} c^{\gamma}_{~\dot{\alpha}} \hspace{0.2cm} =
  \hspace{0.2cm} - c_{\dot{\alpha}}^{~\gamma} \left( \Sigma_{\mu}
  \right)_{\gamma}^{~\dot{\beta}},
   \nn \\
\left( \Sigma_{\mu} \right)_{\dot{\alpha}}^{~\dot{\beta}} & = & \left( \Sigma_{\mu}
 \right)_{\dot{\alpha}}^{~\gamma} c_{\gamma}^{~\dot{\beta}}. \label{sigmasecondset}
\enn

We now define supercoherent states associated with the ULWR in the non-compact basis as,
\eq
e^{( i x^{\mu} P_{\mu} + i \overline{\theta}^{A I} Q_{A I} )}
 | \zeta \rangle = e^{( i x^{\mu} P_{\mu} )} e^{( i \overline{\theta}^{A I} Q_{A I} )}
  |\zeta \rangle, \en where $| \zeta \rangle$ is the chiral primary state and $\theta_{A I}$
are the Grassmann variables obeying, \eq \overline{\theta}^{A I} \hspace{0.2cm} =
\hspace{0.2cm} \theta^{B I} \left( \Gamma_{0} \right)_{B}^{~A} \hspace{0.2cm} \equiv
\hspace{0.2cm} \left( \theta_{B I} \right)^{\dag} \left( \Gamma_{0} \right)_{B}^{~A}. \en

Since $Q_{A I}$ is a right-handed spinor, $\theta_{A I}$ is chosen as a left-handed
spinor, so that $\overline{\theta}^{A I} Q_{A I}$ is a Lorentz singlet : \eq \theta_{A I}
= (\Gamma_{7})_{A}^{~B} \theta_{B I}. \label{thetatheta} \en

If one denotes the hermitian
conjugate of the supersymmetry generators $Q_{A I}$ as \eq Q^{A I} \equiv \left( Q_{A I}
\right)^{\dag}, \en the Dirac conjugation can be written as \eqn
\overline{Q}^{A I} & \equiv & Q^{B I} \left( \Gamma_{0} \right)_{B}^{~A} \nn \\
& = & C^{A B} \Omega^{IJ} Q_{B J}. \enn

To make $e^{i \overline{\theta} Q}$ unitary, we
require its  argument $\overline{\theta}^{A I} Q_{A I}$ to be hermitian, which imposes
the following constraint on $\theta_{A I}$ : \eq \overline{\theta}^{A I} = C^{A B}
\Omega^{IJ} \theta_{B J}. \label{thetabartheta} \en

Explicitly, $\overline{\theta}^{A I}
Q_{A I}$ is given by
\eqn
\overline{\theta}^{A I} Q_{A I} & =  & 2 \left( \overline{\theta}^{1 I} Q_{1 I} +
\overline{\theta}^{2 I} Q_{2 I} + \overline{\theta}^{3 I} Q_{3 I} +
\overline{\theta}^{4 I} Q_{4 I} \right) \nn \\
& = & \frac{i}{\sqrt{2}} \left\{ \overline{\theta}^{1 I} \left( - \overline{s}^{3} \Lambda_{I}
- s^{3} \Lambda^{J} \Omega_{JI} \right) + \overline{\theta}^{2 I} \left( \overline{s}^{4}
\Lambda_{I} + s^{4} \Lambda^{J} \Omega_{JI} \right) \right. \nn \\
& & \hspace{0.75cm} \left. + \overline{\theta}^{3 I} \left( \overline{s}^{1} \Lambda_{I}
+ s^{1} \Lambda^{J} \Omega_{JI} \right) + \overline{\theta}^{4 I} \left( - \overline{s}^{2}
\Lambda_{I} - s^{2} \Lambda^{J} \Omega_{JI} \right) \right\} .
\enn

Then using equations (\ref{thetabartheta}) and (\ref{thetatheta}), one obtains
\eqn
\overline{\theta}^{A I} Q_{A I} & = & - \frac{i}{\sqrt{2}} \left\{ \theta_{1 I}
 \left( s^{1} \Lambda^{I} + \overline{s}^{1} \Omega^{IJ} \Lambda_{J} \right)
 + \theta_{2 I} \left( s^{2} \Lambda^{I} + \overline{s}^{2} \Omega^{IJ} \Lambda_{J} \right)
  \right. \nn \\
& & \hspace{0.75cm} \left. + \theta_{3 I} \left( s^{3} \Lambda^{I} + \overline{s}^{3}
\Omega^{IJ} \Lambda_{J} \right) + \theta_{4 I} \left( s^{4} \Lambda^{I} + \overline{s}^{4}
 \Omega^{IJ} \Lambda_{J} \right) \right\} .
\enn

Note that the index $A$ in $\theta_{A I}$ in the above expression goes from 1 to 4 and
 therefore from this point onwards, we denote it as $\alpha = 1,2,3,4$ ($SU^{*}(4)$ indices)
 and thus write the Grassmann variables as $\theta_{\alpha I}$. Therefore, we have
(cf. equation (\ref{qsinpsi}))
\eq
\overline{\theta}^{A I} Q_{A I} = - \frac{i}{\sqrt{2}} \theta_{\alpha I}
 \left( \overline{s}^{\alpha} \Omega^{IJ} \Lambda_{J} + s^{\alpha} \Lambda^{I} \right).
\en

We restrict ourselves in the rest of the paper to the case $N = 2$. For working in the
 non-compact (5-graded) basis, we define a new Fock vacuum as
\eq
| \tilde{0} \rangle = \left\{ \begin{array}{ll} \beta^{1} \beta^{2} U
| 0 \rangle & \mbox{for $P = 1$,} \\
\beta^{1}(1) \beta^{2}(1) \beta^{1}(2) \beta^{2}(2) U | 0 \rangle & \mbox{for $P = 2$,}
\end{array} \right.
\en
where $| 0 \rangle$ is the ``ordinary'' Fock vacuum, annihilated by $a_{i}$, $b_{j}$,
 $\alpha_{\kappa}$ and $\beta_{\lambda}$. Note that $\Lambda_{I} | \tilde{0} \rangle = 0$.

We also use the notation $\Lambda^{[ K} \Lambda^{L ]|}$ to denote  anti-symmetric symplectic
traceless tensors :
\eq
\Lambda^{[ K} \Lambda^{L ]|} = \Lambda^{[ K} \Lambda^{L ]} - \frac{1}{4} \Omega^{KL}
 \Omega_{K^{\prime} L^{\prime}} \Lambda^{K^{\prime}} \Lambda^{L^{\prime}}.
\en

In Appendix D, we introduce the compact $Spin(5) ( \cong USp(4))$ gamma matrices
 $\gamma_{X}$ ($X = 1,2,\ldots,5$), which are $4 \times 4$ matrices with spinor
 indices $( \gamma_{X} )_{IJ}$, where $I,J = 1,2,3,4$. These matrices are useful
 in projecting out the irreducible $Spin(5)$ representations in the Fock space of
 the fermionic oscillators $\Lambda_{I}$, $\Lambda^{J}$. e.g.
   $( \gamma_{X} )_{KL} \Lambda^{K} \Lambda^{L}$ transforms in the $\underline{5}$ of $Spin(5)$
    or $USp(4)$.

\subsection{Doubleton ($P = 1$) supercoherent states and the corresponding  massless
conformal superfields in $d = 6$}

The supercoherent state determined  by the chiral
primary state $| \zeta \rangle = ( \gamma_{X} )_{KL} \Lambda^{K} \Lambda^{L} | \tilde{0}
\rangle$ is \eqn
e^{i x^{\mu} P_{\mu}} e^{i \overline{\theta} Q} ( \gamma_{X} )_{KL} \Lambda^{K} \Lambda^{L}
| \tilde{0} \rangle & = & ( \gamma_{X} )_{KL} \Lambda^{K} \Lambda^{L} e^{i x^{\mu} P_{\mu}}
| \tilde{0} \rangle \nn \\
& & - \frac{1}{\sqrt{2}} \theta_{\alpha I} \left\{ 2 \overline{s}^{\alpha}
( \gamma_{X} )_{K}^{~I} \Lambda^{K} \right. \nn \\
& & \hspace{1.5cm} \left. - s^{\alpha} ( \gamma_{X} )_{KL} \Lambda^{I}
 \Lambda^{K} \Lambda^{L} \right\} e^{i x^{\mu} P_{\mu}} | \tilde{0} \rangle \nn \\
& & + \frac{1}{4} \theta_{\alpha I} \theta_{\beta J} \left\{ 2 \overline{s}^{( \alpha}
 \overline{s}^{\alpha )} ( \gamma_{X} )^{IJ} \right. \nn \\
& & \hspace{1.8cm} + s^{( \alpha} s^{\beta )} ( \gamma_{X} )_{K L} \Lambda^{I} \Lambda^{J}
 \Lambda^{K} \Lambda^{L} \nn \\
& & \hspace{1.8cm} \left. - \overline{s}^{( \alpha} s^{\beta )}
\left( 4 (\gamma_{X} )_{K}^{~[ I} \Lambda^{J ]} \Lambda^{K} +
\Omega^{I J} ( \gamma_{X} )_{K L} \Lambda^{K} \Lambda^{L} \right) \right\} e^{i x^{\mu} P_{\mu}} | \tilde{0} \rangle \nn \\
& & - \theta_{\alpha I} \theta_{\beta J} \overline{s}^{[ \alpha} s^{\beta ]} (\gamma_{X} )_{K}^{~( I} \Lambda^{J )} \Lambda^{K} e^{i x^{\mu} P_{\mu}} | \tilde{0} \rangle \nn \\
& & + \mathcal{O} ( \theta \theta \theta ).
\enn

We identify $\overline{s}^{[ \alpha} s^{\beta ]} (\gamma_{X} )_{K}^{~( I} \Lambda^{J )}
\Lambda^{K} e^{i x^{\mu} P_{\mu}} | \tilde{0} \rangle$ and the higher order terms $\mathcal{O}
( \theta \theta \theta )$ in the above expansion as ``derivative terms''
 (``excitations'' in the language of particle basis) as they are not annihilated by
  the special conformal generators $K_{\mu}$. On the other hand, the terms which are
   annihilated by $K_{\mu}$ correspond to the component fields of the CPT self-conjugate
    doubleton supermultiplet.

Thus one can write
\eqn
e^{i x^{\mu} P_{\mu}} e^{i \overline{\theta} Q} ( \gamma_{X} )_{KL} \Lambda^{K} \Lambda^{L}
| \tilde{0} \rangle & \cong & \left[ \mbox{(0,0,0)}_{D} , \underline{5} \right]
+ \left[ \mbox{(1,0,0)}_{D} , \underline{4} \right] + \left[ \mbox{(2,0,0)}_{D} ,
 \underline{1} \right] \nn \\
& & + \mbox{derivative terms.}
\enn

This supercoherent state corresponds to Table 1 in \cite{mgst}, which is given below.
  We draw here the correspondence between the states in the compact $SU(4) \times SU(2)$
  basis and the non-compact $SU^{*}(4) \times USp(4)$ basis.
\vspace{.2cm}
\begin{center}
\begin{tabular}{|c|c|c|}
\hline
$SU(4) \times SU(2)$ & $SU^{*}(4) \times USp(4)$ & $USp(4)$ \\ \hline \hline
$| 0 \rangle$ & $( \gamma_{X} )_{KL} \Lambda^{K} \Lambda^{L} e^{i x^{\mu} P_{\mu}}
| \tilde{0} \rangle$ & 5 \\ \hline
$| \onebox , \onebox \rangle$ & $\overline{s}^{\alpha} ( \gamma_{X} )_{K}^{~I}
 \Lambda^{K} e^{i x^{\mu} P_{\mu}} | \tilde{0} \rangle$ & 4 \\
& $s^{\alpha} ( \gamma_{X} )_{KL} \Lambda^{I} \Lambda^{K}
\Lambda^{L} e^{i x^{\mu} P_{\mu}} | \tilde{0} \rangle$ & \\ \hline
$| \twobox , \oneonebox \rangle$ & $\overline{s}^{( \alpha} \overline{s}^{\beta )}
 ( \gamma_{X} )^{IJ} e^{i x^{\mu} P_{\mu}} | \tilde{0} \rangle$ & 1 \\
& $s^{( \alpha} s^{\beta )} ( \gamma_{X} )_{KL} \Lambda^{I} \Lambda^{J} \Lambda^{K}
\Lambda^{L} e^{i x^{\mu} P_{\mu}} | \tilde{0} \rangle$ & \\
& $\overline{s}^{( \alpha} s^{\beta )} \left( 4 ( \gamma_{X} )_{K}^{~[ I} \Lambda^{J ]}
 \Lambda^{K} + \Omega^{IJ} (\gamma_{X} )_{KL} \Lambda^{K} \Lambda^{L}
 \right) e^{i x^{\mu} P_{\mu}} | \tilde{0} \rangle$ & \\ \hline
\end{tabular}
\end{center}
\begin{center} Table 3. The CPT self-conjugate doubleton supermultiplet in
the supercoherent state basis. \end{center} \vspace{.2cm}

Similarly, the supercoherent
state determined by the chiral  primary state $| \zeta \rangle = \Lambda^{L} | \tilde{0}
\rangle$ is \eqn
e^{i x^{\mu} P_{\mu}} e^{i \overline{\theta} Q} \Lambda^{L} | \tilde{0} \rangle & =
 & \Lambda^{L} e^{i x^{\mu} P_{\mu}} | \tilde{0} \rangle \nn \\
& & + \frac{1}{\sqrt{2}} \theta_{\alpha I} s^{\alpha}
\Lambda^{[ I} \Lambda^{L ]|} e^{i x^{\mu} P_{\mu}} | \tilde{0} \rangle \nn \\
& & + \frac{1}{\sqrt{2}} \theta_{\alpha I} \left\{ \overline{s}^{\alpha}
\Omega^{I L} \right. \nn \\
& & \hspace{1.5cm} \left. + \frac{1}{4} s^{\alpha} \Omega^{I L} \Omega_{I^{\prime} L^{\prime}}
 \Lambda^{I^{\prime}} \Lambda^{L^{\prime}} \right\} e^{i x^{\mu} P_{\mu}}
 | \tilde{0} \rangle \nn \\
& & - \frac{1}{4} \theta_{\alpha I} \theta_{\beta J} \left\{ s^{( \alpha} s^{\beta )}
 \Lambda^{[ I} \Lambda^{J} \Lambda^{L ]} \right. \nn \\
& & \hspace{1.8cm} \left. + \overline{s}^{( \alpha} s^{\beta )} \left( \Omega^{I J}
 \Lambda^{L} + 2 \Omega^{L [ I} \Lambda^{J ]} \right) \right\} e^{i x^{\mu} P_{\mu}}
 | \tilde{0} \rangle \nn \\
& & - \frac{1}{12\sqrt{2}} \theta_{\alpha I} \theta_{\beta J} \theta_{\gamma K}
 s^{( \alpha} s^{\beta} s^{\gamma )} \Lambda^{[ I} \Lambda^{J} \Lambda^{K} \Lambda^{L ]}
  e^{i x^{\mu} P_{\mu}} | \tilde{0} \rangle \nn \\
& & - \frac{1}{2} \theta_{\alpha I} \theta_{\beta J} \overline{s}^{[ \alpha} s^{\beta ]}
 \Omega^{L ( I} \Lambda^{J )} e^{i x^{\mu} P_{\mu}} | \tilde{0} \rangle \nn \\
& & - \frac{1}{12\sqrt{2}} \theta_{\alpha I} \theta_{\beta J} \theta_{\gamma K}
\left\{ \overline{s}^{\alpha} \overline{s}^{\beta} s^{\gamma}
\left( \Omega^{I L} \Omega^{J K} - 2 \Omega^{I K} \Omega^{J L} \right) \right. \nn \\
& & \hspace{3.2cm} \left. + \overline{s}^{\alpha} s^{\beta} s^{\gamma}
\left( \Omega^{I J} \Lambda^{K} \Lambda^{L} - 2 \Omega^{I K} \Lambda^{J} \Lambda^{L} + 3
\Omega^{I L} \Lambda^{J} \Lambda^{K} \right) \right\} e^{i x^{\mu} P_{\mu}}
 | \tilde{0} \rangle \nn \\
& & + \mathcal{O} ( \theta \theta \theta \theta ) \nn \\
& \cong & \left[ \mbox{(0,0,0)}_{D} , \underline{4} \right] + \left[ \mbox{(1,0,0)}_{D} ,
 \underline{5} \right] + \left[ \mbox{(1,0,0)}_{D} , \underline{1} \right] +
 \left[ \mbox{(2,0,0)}_{D} , \underline{4} \right] + \left[ \mbox{(3,0,0)}_{D} ,
 \underline{1} \right] \nn \\
& & + \mbox{derivative terms.}
\enn

Above, $\left( \overline{s}^{[ \alpha} s^{\beta ]} \ldots \right)$,
$\left( \overline{s}^{\alpha} \overline{s}^{\beta} s^{\gamma} \ldots \right)$,
 $\left( \overline{s}^{\alpha} s^{\beta} s^{\gamma} \ldots \right)$ and $\mathcal{O}
 ( \theta \theta \theta \theta )$ are derivative terms.

This supercoherent state corresponds to Table 2 in \cite{mgst}, which was
generated by the lowest weight vector $ | \Omega \rangle = | \sonebox \rangle$
in the compact basis.
\vspace{.2cm}
\begin{center}
\begin{tabular}{|c|c|c|}
\hline
$SU(4) \times SU(2)$ & $SU^{*}(4) \times USp(4)$ & $USp(4)$ \\ \hline \hline
$| 1 , \onebox \rangle$ & $\Lambda^{L} e^{i x^{\mu} P_{\mu}} | \tilde{0} \rangle$ & 4 \\ \hline
$| \onebox , 1 \rangle$ & $s^{\alpha} \Lambda^{[ I} \Lambda^{L ]|} e^{i x^{\mu} P_{\mu}}
| \tilde{0} \rangle$ & 5 \\ \hline
$| \onebox , \oneonebox \rangle$ & $\overline{s}^{\alpha} \Omega^{I L} e^{i x^{\mu}
P_{\mu}} | \tilde{0} \rangle$ & 1 \\
& $s^{\alpha} \Omega^{I L} \Omega_{I^{\prime} L^{\prime}} \Lambda^{I^{\prime}}
 \Lambda^{L^{\prime}} e^{i x^{\mu} P_{\mu}} | \tilde{0} \rangle$ & \\ \hline
$| \twobox , \onebox \rangle$ & $s^{( \alpha} s^{\beta )} \Lambda^{[ I} \Lambda^{J}
\Lambda^{L ]} e^{i x^{\mu} P_{\mu}} | \tilde{0} \rangle$ & 4 \\
& $\overline{s}^{( \alpha} s^{\beta )} \left( \Omega^{I J} \Lambda^{L} +
2 \Omega^{L [ I} \Lambda^{J ]} \right) e^{i x^{\mu} P_{\mu}} | \tilde{0} \rangle$ & \\ \hline
$| \threebox , \oneonebox \rangle$ & $s^{( \alpha} s^{\beta} s^{\gamma )}
 \Lambda^{[ I} \Lambda^{J} \Lambda^{K} \Lambda^{L ]} e^{i x^{\mu} P_{\mu}}
 | \tilde{0} \rangle$ & 1 \\ \hline
\end{tabular}
\end{center}
\begin{center} Table 4. The doubleton supermultiplet  defined by the chiral primary state
 $| \zeta \rangle = \Lambda^{L} | \tilde{0} \rangle$ in the supercoherent state basis.
 \end{center}

The supercoherent state determined by the chiral primary state $| \zeta \rangle =
| \tilde{0} \rangle$, which corresponds to Table 3 in \cite{mgst} (for $j = 1$) is
\eqn
e^{i x^{\mu} P_{\mu}} e^{i \overline{\theta} Q} | \tilde{0} \rangle & = & e^{i x^{\mu}
 P_{\mu}} | \tilde{0} \rangle \nn \\
& & + \frac{1}{\sqrt{2}} \theta_{\alpha I} s^{\alpha} \Lambda^{I} e^{i x^{\mu} P_{\mu}}
| \tilde{0} \rangle \nn \\
& & - \frac{1}{4} \theta_{\alpha I} \theta_{\beta J} \left\{ \frac{1}{4} s^{( \alpha}
 s^{\beta )} \Omega^{I J} \Omega_{I^{\prime} J^{\prime}} \Lambda^{I^{\prime}}
 \Lambda^{J^{\prime}} \right. \nn \\
& & \hspace{1.8cm} \left. + \overline{s}^{( \alpha} s^{\beta )} \Omega^{I J} \right\}
e^{i x^{\mu} P_{\mu}} | \tilde{0} \rangle \nn \\
& & - \frac{1}{4} \theta_{\alpha I} \theta_{\beta J} s^{( \alpha} s^{\beta )}
\Lambda^{[ I} \Lambda^{J ]|} e^{i x^{\mu} P_{\mu}} | \tilde{0} \rangle \nn \\
& & - \frac{1}{12\sqrt{2}} \theta_{\alpha I} \theta_{\beta J} \theta_{\gamma K}
 s^{( \alpha} s^{\beta} s^{\gamma )} \Lambda^{[ I} \Lambda^{J} \Lambda^{K ]}
  e^{i x^{\mu} P_{\mu}} | \tilde{0} \rangle \nn \\
& & + \frac{1}{96} \theta_{\alpha I} \theta_{\beta J} \theta_{\gamma K}
\theta_{\delta L} s^{( \alpha} s^{\beta} s^{\gamma} s^{\delta )} \Lambda^{[ I} \Lambda^{J}
 \Lambda^{K} \Lambda^{L ]} e^{i x^{\mu} P_{\mu}} | \tilde{0} \rangle \nn \\
& & - \frac{1}{12\sqrt{2}} \theta_{\alpha I} \theta_{\beta J} \theta_{\gamma K}
\overline{s}^{\alpha} s^{\beta} s^{\gamma} \left( \Omega^{I J} \Lambda^{K} - 2 \Omega^{I K}
 \Lambda^{J} \right) e^{i x^{\mu} P_{\mu}} | \tilde{0} \rangle \nn \\
& & + \frac{1}{96} \theta_{\alpha I} \theta_{\beta J} \theta_{\gamma K}
\theta_{\delta L} \left\{ \overline{s}^{\alpha} \overline{s}^{\beta} s^{\gamma}
 s^{\delta} \left( \Omega^{I L} \Omega^{J K} - 2 \Omega^{I K} \Omega^{J L} \right)
 \right. \nn \\
& & \hspace{3.1cm} \left. + \overline{s}^{\alpha} s^{\beta} s^{\gamma} s^{\delta}
\left( \Omega^{I J} \Lambda^{K} \Lambda^{L} - 2 \Omega^{I K} \Lambda^{J} \Lambda^{L} +
3 \Omega^{I L} \Lambda^{J} \Lambda^{K} \right) \right\} e^{i x^{\mu} P_{\mu}}
| \tilde{0}\rangle \nn \\
& & + \mathcal{O} ( \theta \theta \theta \theta \theta ) \nn \\
& \cong & \left[ \mbox{(0,0,0)}_{D},\underline{1} \right] +
 \left[ \mbox{(1,0,0)}_{D},\underline{4} \right] +
  \left[ \mbox{(2,0,0)}_{D},\underline{1} \right] +
  \left[ \mbox{(2,0,0)}_{D},\underline{5} \right] +
  \left[ \mbox{(3,0,0)}_{D},\underline{4} \right] \nn \\
& & + \left[ \mbox{(4,0,0)}_{D},\underline{1} \right] + \mbox{derivative terms.}
\enn

In this expression, $\left( \overline{s}^{\alpha} s^{\beta} s^{\gamma}
 \ldots \right)$, $\left( \overline{s}^{\alpha} \overline{s}^{\beta} s^{\gamma}
  s^{\delta} \ldots \right)$, $\left( \overline{s}^{\alpha} s^{\beta} s^{\gamma}
   s^{\delta} \ldots \right)$ and $\mathcal{O} ( \theta \theta \theta \theta \theta )$
   are derivative terms.
\vspace{.2cm}
\begin{center}
\begin{tabular}{|c|c|c|}
\hline
$SU(4) \times SU(2)$ & $SU^{*}(4) \times USp(4)$ & $USp(4)$ \\ \hline \hline
$| 1 , \oneonebox \rangle$ & $e^{i x^{\mu} P_{\mu}} | \tilde{0} \rangle$ & 1 \\ \hline
$| \onebox , \onebox \rangle$ & $s^{\alpha} \Lambda^{I} e^{i x^{\mu} P_{\mu}} | \tilde{0}
 \rangle$ & 4 \\ \hline
$| \twobox , \oneonebox \rangle$ & $s^{( \alpha} s^{\beta )} \Omega^{I J} \Omega_{I^{\prime}
 J^{\prime}} \Lambda^{I^{\prime}} \Lambda^{J^{\prime}} e^{i x^{\mu} P_{\mu}} | \tilde{0}
  \rangle$ & 1 \\
& $\overline{s}^{( \alpha} s^{\beta )} \Omega^{I J} e^{i x^{\mu} P_{\mu}} | \tilde{0}
\rangle$ & \\ \hline
$| \twobox , 1 \rangle$ & $s^{( \alpha} s^{\beta )} \Lambda^{[ I} \Lambda^{J ]|}
e^{i x^{\mu} P_{\mu}} | \tilde{0} \rangle$ & 5 \\ \hline
$| \threebox , \onebox \rangle$ & $s^{( \alpha} s^{\beta} s^{\gamma )} \Lambda^{[ I}
\Lambda^{J} \Lambda^{K ]} e^{i x^{\mu} P_{\mu}} | \tilde{0} \rangle$ & 4 \\ \hline
$| \fourbox , \oneonebox \rangle$ & $s^{( \alpha} s^{\beta} s^{\gamma} s^{\delta )}
\Lambda^{[ I} \Lambda^{J} \Lambda^{K} \Lambda^{L ]} e^{i x^{\mu} P_{\mu}}
| \tilde{0} \rangle$ & 1 \\ \hline
\end{tabular}
\end{center}
\begin{center} Table 5. The doubleton supermultiplet defined by
the chiral primary state $| \zeta \rangle = | \tilde{0} \rangle$  in the supercoherent
state basis. \end{center} \vspace{.2cm}

Following the same procedure, we give below the
supercoherent state obtained, starting from the chiral primary state $| \zeta \rangle =
s^{\eta} | \tilde{0} \rangle$, which corresponds to Table 3 in \cite{mgst} (for $j =
\frac{3}{2}$). \eqn
e^{i x^{\mu} P_{\mu}} e^{i \overline{\theta} Q} s^{\eta} | \tilde{0} \rangle & =
 & s^{\eta} e^{i x^{\mu} P_{\mu}} | \tilde{0} \rangle \nn \\
& & + \frac{1}{\sqrt{2}} \theta_{\alpha I} s^{( \alpha} s^{\eta )} \Lambda^{I}
 e^{i x^{\mu} P_{\mu}} | \tilde{0} \rangle \nn \\
& & - \frac{1}{16} \theta_{\alpha I} \theta_{\beta J} s^{( \alpha} s^{\beta} s^{\eta )}
\Omega^{I J} \Omega_{I^{\prime} J^{\prime}} \Lambda^{I^{\prime}} \Lambda^{J^{\prime}}
 e^{i x^{\mu} P_{\mu}} | \tilde{0} \rangle \nn \\
& & - \frac{1}{4} \theta_{\alpha I} \theta_{\beta J} s^{( \alpha} s^{\beta} s^{\eta )}
 \Lambda^{[ I} \Lambda^{J ]|} e^{i x^{\mu} P_{\mu}} | \tilde{0} \rangle \nn \\
& & - \frac{1}{12\sqrt{2}} \theta_{\alpha I} \theta_{\beta J} \theta_{\gamma K}
s^{( \alpha} s^{\beta} s^{\gamma} s^{\eta )} \Lambda^{[ I} \Lambda^{J} \Lambda^{K ]}
 e^{i x^{\mu} P_{\mu}} | \tilde{0} \rangle \nn \\
& & + \frac{1}{96} \theta_{\alpha I} \theta_{\beta J} \theta_{\gamma K}
\theta_{\delta L} s^{( \alpha} s^{\beta} s^{\gamma} s^{\delta} s^{\eta )}
\Lambda^{[ I} \Lambda^{J} \Lambda^{K} \Lambda^{L ]} e^{i x^{\mu} P_{\mu}}
| \tilde{0} \rangle \nn \\
& & - \frac{1}{4} \theta_{\alpha I} \theta_{\beta J} \overline{s}^{\alpha}
 s^{\beta} s^{\eta} \Omega^{I J} e^{i x^{\mu} P_{\mu}} | \tilde{0} \rangle \nn \\
& & - \frac{1}{12\sqrt{2}} \theta_{\alpha I} \theta_{\beta J} \theta_{\gamma K}
 \overline{s}^{\alpha} s^{\beta} s^{\gamma} s^{\eta} \left( \Omega^{I J} \Lambda^{K} - 2
  \Omega^{I K} \Lambda^{J} \right) e^{i x^{\mu} P_{\mu}} | \tilde{0} \rangle \nn \\
& & + \frac{1}{96} \theta_{\alpha I} \theta_{\beta J} \theta_{\gamma K} \theta_{\delta L}
 \left\{ \overline{s}^{\alpha} \overline{s}^{\beta} s^{\gamma} s^{\delta} s^{\eta}
 \left( \Omega^{I L} \Omega^{J K} - 2 \Omega^{I K} \Omega^{J L} \right) \right. \nn \\
& & \hspace{3.2cm} \left. + \overline{s}^{\alpha} s^{\beta} s^{\gamma} s^{\delta} s^{\eta}
\left( \Omega^{I J} \Lambda^{K} \Lambda^{L} - 2 \Omega^{I K} \Lambda^{J} \Lambda^{L} +
 3 \Omega^{I L} \Lambda^{J} \Lambda^{K} \right) \right\} e^{i x^{\mu} P_{\mu}} | \tilde{0}
 \rangle \nn \\
& & + \mathcal{O} ( \theta \theta \theta \theta \theta ) \nn \\
& \cong & \left[ \mbox{(1,0,0)}_{D},\underline{1} \right] + \left[ \mbox{(2,0,0)}_{D},
\underline{4} \right] + \left[ \mbox{(3,0,0)}_{D},\underline{1} \right] +
\left[ \mbox{(3,0,0)}_{D},\underline{5} \right] + \left[ \mbox{(4,0,0)}_{D},\underline{4}
\right] \nn \\
& & + \left[ \mbox{(5,0,0)}_{D},\underline{1} \right] + \mbox{derivative terms.}
\enn

The derivative terms are, $\left( \overline{s}^{\alpha} s^{\beta} s^{\eta} \ldots
\right)$, $\left( \overline{s}^{\alpha} s^{\beta} s^{\gamma} s^{\eta} \ldots \right)$,
 $\left( \overline{s}^{\alpha} \overline{s}^{\beta} s^{\gamma} s^{\delta} s^{\eta} \ldots
 \right)$, $\left( \overline{s}^{\alpha} s^{\beta} s^{\gamma} s^{\delta} s^{\eta} \ldots
 \right)$ and $\mathcal{O} ( \theta \theta \theta \theta \theta )$.
\vspace{.2cm}
\begin{center}
\begin{tabular}{|c|c|c|}
\hline
$SU(4) \times SU(2)$ & $SU^{*}(4) \times USp(4)$ & $USp(4)$ \\ \hline \hline
$| \onebox , \oneonebox \rangle$ & $s^{\eta} e^{i x^{\mu} P_{\mu}} | \tilde{0} \rangle$ & 1 \\
 \hline
$| \twobox , \onebox \rangle$ & $s^{( \alpha} s^{\eta )} \Lambda^{I} e^{i x^{\mu} P_{\mu}}
 | \tilde{0} \rangle$ & 4 \\ \hline
$| \threebox , 1 \rangle$ & $s^{( \alpha} s^{\beta} s^{\eta )} \Omega^{I J} \Omega_{I^{\prime}
 J^{\prime}} \Lambda^{I^{\prime}} \Lambda^{J^{\prime}} e^{i x^{\mu} P_{\mu}}
  | \tilde{0} \rangle$ & 1 \\ \hline
$| \threebox , \oneonebox \rangle$ & $s^{( \alpha} s^{\beta} s^{\eta )}
\Lambda^{[ I} \Lambda^{J ]|} e^{i x^{\mu} P_{\mu}} | \tilde{0} \rangle$ & 5 \\ \hline
$| \fourbox , \onebox \rangle$ & $ s^{( \alpha} s^{\beta} s^{\gamma} s^{\eta )}
 \Lambda^{[ I} \Lambda^{J} \Lambda^{K ]} e^{i x^{\mu} P_{\mu}} | \tilde{0} \rangle$ & 4 \\ \hline
$| \fivebox , \oneonebox \rangle$ & $s^{( \alpha} s^{\beta} s^{\gamma} s^{\delta} s^{\eta )}
\Lambda^{[ I} \Lambda^{J} \Lambda^{K} \Lambda^{L ]} e^{i x^{\mu} P_{\mu}} | \tilde{0} \rangle$
 & 1 \\ \hline
\end{tabular}
\end{center}
\begin{center} Table 6. The doubleton supermultiplet defined by   the chiral primary state
 $| \zeta \rangle = s^{\eta} | \tilde{0} \rangle$ in the supercoherent state basis. \end{center}

Finally we give the supercoherent state corresponding to the general  doubleton
supermultiplet, which is determined by the chiral primary state $| \zeta \rangle = s^{(
\beta_{1}} \ldots s^{\beta_{n} )} | \tilde{0} \rangle$. In \cite{mgst}, this
supermultiplet was obtained in the compact basis from the lowest weight vector $| \Omega
\rangle = | \smarctwojbox \rangle$ where $j > \frac{1}{2}$. \eqn
\lefteqn{e^{i x^{\mu} P_{\mu}} e^{i \overline{\theta} Q} s^{( \beta_{1}} \ldots s^{\beta_{n} )}
| \tilde{0} \rangle = } \nn \\
& & s^{( \beta_{1}} \ldots s^{\beta_{n} )} e^{i x^{\mu} P_{\mu}} | \tilde{0} \rangle \nn \\
& & + \frac{1}{\sqrt{2}} \theta_{\alpha_{1} I_{1}} s^{( \alpha_{1}} s^{\beta_{1}}
 \ldots s^{\beta_{n} )} \Lambda^{I_{1}} e^{i x^{\mu} P_{\mu}} | \tilde{0} \rangle \nn \\
& & - \frac{1}{4} \theta_{\alpha_{1} I_{1}} \theta_{\alpha_{2} I_{2}} s^{( \alpha_{1}}
 s^{\alpha_{2}} s^{\beta_{1}} \ldots s^{\beta_{n} )} \Lambda^{[ I_{1}} \Lambda^{I_{2} ]|}
 e^{i x^{\mu} P_{\mu}} | \tilde{0} \rangle \nn \\
& & - \frac{1}{16} \theta_{\alpha_{1} I_{1}} \theta_{\alpha_{2} I_{2}} s^{( \alpha_{1}}
s^{\alpha_{2}} s^{\beta_{1}} \ldots s^{\beta_{n} )} \Omega^{I_{1} I_{2}}
\Omega_{I_{1}^{\prime} I_{2}^{\prime}} \Lambda^{I_{1}^{\prime}} \Lambda^{I_{2}^{\prime}}
 e^{i x^{\mu} P_{\mu}} | \tilde{0} \rangle \nn \\
& & - \frac{1}{12\sqrt{2}} \theta_{\alpha_{1} I_{1}} \theta_{\alpha_{2} I_{2}}
\theta_{\alpha_{3} I_{3}} s^{( \alpha_{1}} s^{\alpha_{2}} s^{\alpha_{3}} s^{\beta_{1}}
 \ldots s^{\beta_{n} )} \Lambda^{[ I_{1}} \Lambda^{I_{2}} \Lambda^{I_{3} ]} e^{i x^{\mu}
 P_{\mu}} | \tilde{0} \rangle \nn \\
& & + \frac{1}{96} \theta_{\alpha_{1} I_{1}} \theta_{\alpha_{2} I_{2}} \theta_{\alpha_{3}
I_{3}} \theta_{\alpha_{4} I_{4}} s^{( \alpha_{1}} s^{\alpha_{2}} s^{\alpha_{3}} s^{\alpha_{4}}
 s^{\beta_{1}} \ldots s^{\beta_{n} )} \Lambda^{[ I_{1}} \Lambda^{I_{2}} \Lambda^{I_{3}}
 \Lambda^{I_{4} ]} e^{i x^{\mu} P_{\mu}} | \tilde{0} \rangle \nn \\
& & - \frac{1}{4} \theta_{\alpha_{1} I_{1}} \theta_{\alpha_{2} I_{2}}
 \overline{s}^{\alpha_{1}} s^{\alpha_{2}} s^{\beta_{1}} \ldots s^{\beta_{n}}
 \Omega^{I_{1} I_{2}} e^{i x^{\mu} P_{\mu}} | \tilde{0} \rangle \nn \\
& & - \frac{1}{12\sqrt{2}} \theta_{\alpha_{1} I_{1}} \theta_{\alpha_{2} I_{2}}
 \theta_{\alpha_{3} I_{3}} \overline{s}^{\alpha_{1}} s^{\alpha_{2}} s^{\alpha_{3}}
 s^{\beta_{1}} \ldots s^{\beta_{n}} \left( \Omega^{I_{1} I_{2}} \Lambda^{I_{3}} - 2
 \Omega^{I_{1} I_{3}} \Lambda^{I_{2}} \right) e^{i x^{\mu} P_{\mu}}
  | \tilde{0} \rangle \nn \\
& & + \frac{1}{96} \theta_{\alpha_{1} I_{1}} \theta_{\alpha_{2} I_{2}}
\theta_{\alpha_{3} I_{3}} \theta_{\alpha_{4} I_{4}} \left\{ \overline{s}^{\alpha_{1}}
\overline{s}^{\alpha_{2}} s^{\alpha_{3}} s^{\alpha_{4}} s^{\beta_{1}} \ldots s^{\beta_{n}}
 \left( \Omega^{I_{1} I_{4}} \Omega^{I_{2} I_{3}} - 2 \Omega^{I_{1} I_{3}} \Omega^{I_{2} I_{4}}
  \right) \right. \nn \\
& & \hspace{4.2cm} - \overline{s}^{\alpha_{1}} s^{\alpha_{2}} s^{\alpha_{3}} s^{\alpha_{4}}
 s^{\beta_{1}} \ldots s^{\beta_{n}} \left( \Omega^{I_{1} I_{2}} \Lambda^{I_{3}} \Lambda^{I_{4}}
  - 2 \Omega^{I_{1} I_{3}} \Lambda^{I_{2}} \Lambda^{I_{4}} \right. \nn \\
& & \hspace{8.5cm} \left. \left. + 3 \Omega^{I_{1} I_{4}} \Lambda^{I_{2}} \Lambda^{I_{3}}
 \right) \right\} e^{i x^{\mu} P_{\mu}} | \tilde{0} \rangle \nn \\
& & + \mathcal{O} ( \theta \theta \theta \theta \theta ) \nn \\
& \cong & \left[ \mbox{(n,0,0)}_{D},\underline{1} \right] + \left[ \mbox{(n+1,0,0)}_{D},
\underline{4} \right] + \left[ \mbox{(n+2,0,0)}_{D},\underline{5} \right] +
\left[ \mbox{(n+2,0,0)}_{D},\underline{1} \right] + \left[ \mbox{(n+3,0,0)}_{D},\underline{4}
 \right] \nn \\
& & + \left[ \mbox{(n+4,0,0)}_{D},\underline{1} \right] + \mbox{derivative terms.}
\enn

We find that the derivative terms of this expression are, $\left( \overline{s}^{\alpha_{1}}
 s^{\alpha_{2}} s^{\beta_{1}} \ldots s^{\beta_{n}} \ldots \ldots \right)$,
 $\left( \overline{s}^{\alpha_{1}} s^{\alpha_{2}} s^{\alpha_{3}}
  s^{\beta_{1}} \ldots s^{\beta_{n}} \ldots \ldots \right)$, $\left( \overline{s}^{\alpha_{1}}
   \overline{s}^{\alpha_{2}} s^{\alpha_{3}} s^{\alpha_{4}} s^{\beta_{1}} \ldots s^{\beta_{n}}
    \ldots \ldots \right)$, $\left( \overline{s}^{\alpha_{1}} s^{\alpha_{2}} s^{\alpha_{3}}
     s^{\alpha_{4}} s^{\beta_{1}} \ldots s^{\beta_{n}} \ldots \ldots \right)$ and $\mathcal{O}
     ( \theta \theta \theta \theta \theta )$.
\vspace{.2cm}
\begin{center}
\begin{tabular}{|c|c|c|}
\hline
$SU(4) \times SU(2)$ & $SU^{*}(4) \times USp(4)$ & $USp(4)$ \\ \hline \hline
$| \marcnbox , \oneonebox \rangle$ & $s^{( \beta_{1}} \ldots s^{\beta_{n} )} e^{i x^{\mu}
P_{\mu}} | \tilde{0} \rangle$ & 1 \\ \hline
$| \marcnplusonebox , \onebox \rangle$ & $s^{( \alpha_{1}} s^{\beta_{1}} \ldots s^{\beta_{n} )}
 \Lambda^{I_{1}} e^{i x^{\mu} P_{\mu}} | \tilde{0} \rangle$ & 4 \\ \hline
$| \marcnplustwobox , 1 \rangle$ & $s^{( \alpha_{1}} s^{\alpha_{2}} s^{\beta_{1}}
 \ldots s^{\beta_{n} )} \Lambda^{[ I_{1}} \Lambda^{I_{2} ]|} e^{i x^{\mu} P_{\mu}} | \tilde{0}
 \rangle$ & 5 \\ \hline
$| \marcnplustwobox , \oneonebox \rangle$ & $s^{( \alpha_{1}} s^{\alpha_{2}} s^{\beta_{1}}
 \ldots s^{\beta_{n} )} \Omega^{I_{1} I_{2}} \Omega_{I_{1}^{\prime} I_{2}^{\prime}}
 \Lambda^{I_{1}^{\prime}} \Lambda^{I_{2}^{\prime}} e^{i x^{\mu} P_{\mu}}
 | \tilde{0}\rangle$ & 1 \\ \hline
$| \marcnplusthreebox , \onebox \rangle$ & $s^{( \alpha_{1}} s^{\alpha_{2}}
 s^{\alpha_{3}} s^{\beta_{1}} \ldots s^{\beta_{n} )} \Lambda^{[ I_{1}}
 \Lambda^{I_{2}} \Lambda^{I_{3} ]} e^{i x^{\mu} P_{\mu}} | \tilde{0} \rangle$ & 4 \\ \hline
$| \marcnplusfourbox , \oneonebox \rangle$ & $s^{( \alpha_{1}} s^{\alpha_{2}}
s^{\alpha_{3}} s^{\alpha_{4}} s^{\beta_{1}} \ldots s^{\beta_{n} )} \Lambda^{[ I_{1}}
 \Lambda^{I_{2}} \Lambda^{I_{3}} \Lambda^{I_{4} ]} e^{i x^{\mu} P_{\mu}} | \tilde{0}
 \rangle$ & 1 \\ \hline
\end{tabular}
\end{center}
\begin{center} Table 7. The general  doubleton supermultiplet defined by the chiral primary
state $| \zeta \rangle = s^{( \beta_{1}} \ldots s^{\beta_{n} )} | \tilde{0} \rangle$  in
the supercoherent state basis. \end{center}

\subsection{Massless $AdS_{7}$ supermultiplets ($P = 2$) versus massive conformal
superfields in $d = 6$}

As has been discussed extensively in  the literature
\cite{gnw,mgst}, the supermultiplets of \OSN ~for $P = 2$ correspond to massless $AdS$
supermultiplets in $d = 7$. However, as conformal superfields in $d = 6$, they are
massive. Hence the massless graviton supermultiplet of \OS ~obtained from the lowest
weight vector $| \Omega \rangle = | 0 \rangle$ for $P = 2$ corresponds to a massive
conformal superfield in $d = 6$. In the 5-graded non-compact basis of \OS, the
corresponding conformal superfield is obtained from the chiral primary state \eq \left(
\mathcal{P}_{X Y} \right)_{I_{1} I_{2} I_{3} I_{4}} \Lambda^{I_{1}}(1) \Lambda^{I_{2}}(1)
\Lambda^{I_{3}}(2) \Lambda^{I_{4}}(2) | \tilde{0} \rangle \en where \eq \left(
\mathcal{P}_{X Y} \right)_{I_{1} I_{2} I_{3} I_{4}} = ( \gamma_{X} )_{I_{1} I_{2}} (
\gamma_{Y} )_{I_{3} I_{4}} - \frac{1}{5} \delta_{X Y} ( \gamma_{Z} )_{I_{1} I_{2}} (
\gamma_{Z} )_{I_{3} I_{4}}. \en

However, below we give the action of $e^{i x^{\mu}
P_{\mu}} e^{i \overline{\theta} Q}$ on $\Lambda^{I_{1}}(1) \Lambda^{I_{2}}(1)
\Lambda^{I_{3}}(2) \Lambda^{I_{4}}(2) | \tilde{0} \rangle$ only. One can project each
term with $\left( \mathcal{P}_{X Y} \right)_{I_{1} I_{2} I_{3} I_{4}}$ to obtain the
component fields of the massless graviton supermultiplet. \eqn
\lefteqn{e^{i x^{\mu} P_{\mu}} e^{i \overline{\theta} Q} \Lambda^{I_{1}}(1)
\Lambda^{I_{2}}(1) \Lambda^{I_{3}}(2) \Lambda^{I_{4}}(2) | \tilde{0} \rangle =}
\nn \\
& & \Lambda^{I_{1}}(1) \Lambda^{I_{2}}(1) \Lambda^{I_{3}}(2) \Lambda^{I_{4}}(2)
 e^{i x^{\mu} P_{\mu}} | \tilde{0} \rangle \nn \\
& & + \sqrt{2} \theta_{\alpha I} \left\{ \overline{s}^{\alpha}(1) \Omega^{I [ I_{1}}
\Lambda^{I_{2} ]}(1) \Lambda^{I_{3}}(2) \Lambda^{I_{4}}(2) \right. \nn \\
& & \hspace{1.5cm} \left. + \overline{s}^{\alpha}(2) \Omega^{I [ I_{3}} \Lambda^{I_{4} ]}(2)
 \Lambda^{I_{1}}(1) \Lambda^{I_{2}}(1) \right\} e^{i x^{\mu} P_{\mu}} | \tilde{0}
 \rangle \nn \\
& & + \frac{1}{\sqrt{2}} \theta_{\alpha I} \left\{ s^{\alpha}(1) \Lambda^{I}(1)
 \Lambda^{I_{1}}(1) \Lambda^{I_{2}}(1) \Lambda^{I_{3}}(2) \Lambda^{I_{4}}(2)
 \right. \nn \\
& & \hspace{1.5cm} \left. + s^{\alpha}(2) \Lambda^{I}(2) \Lambda^{I_{1}}(1)
 \Lambda^{I_{2}}(1) \Lambda^{I_{3}}(2) \Lambda^{I_{4}}(2) \right\} e^{i x^{\mu}
 P_{\mu}} | \tilde{0} \rangle \nn \\
& & - \frac{1}{2} \theta_{\alpha I} \theta_{\beta J} \left\{ \overline{s}^{( \alpha}(1)
\overline{s}^{\beta )}(1) \Omega^{I [ I_{1}} \Omega^{I_{2} ] J} \Lambda^{I_{3}}(2)
\Lambda^{I_{4}}(2) \right. \nn \\
& & \hspace{1.8cm} + \overline{s}^{( \alpha}(2) \overline{s}^{\beta )}(2)
 \Omega^{I [ I_{3}} \Omega^{I_{4} ] J} \Lambda^{I_{1}}(1) \Lambda^{I_{2}}(1) \nn \\
& & \hspace{1.8cm} \left. + 4 \overline{s}^{( \alpha}(1) \overline{s}^{\beta )}(2)
\Lambda^{[ I_{1}}(1) \Omega^{I_{2} ] [ I} \Omega^{J ] [ I_{3}} \Lambda^{I_{4} ]}(2)
 \right\} e^{i x^{\mu} P_{\mu}} | \tilde{0} \rangle \nn \\
& & - \frac{1}{4} \theta_{\alpha I} \theta_{\beta J} \left\{ s^{( \alpha}(1) s^{\beta )}(1)
\Lambda^{I}(1) \Lambda^{J}(1) \Lambda^{I_{1}}(1) \Lambda^{I_{2}}(1) \Lambda^{I_{3}}(2)
 \Lambda^{I_{4}}(2) \right. \nn \\
& & \hspace{1.8cm} + s^{( \alpha}(2) s^{\beta )}(2) \Lambda^{I}(2) \Lambda^{J}(2)
 \Lambda^{I_{1}}(1) \Lambda^{I_{2}}(1) \Lambda^{I_{3}}(2) \Lambda^{I_{4}}(2) \nn \\
& & \hspace{1.8cm} \left. + 2 s^{( \alpha}(1) s^{\beta )}(2) \Lambda^{[ I}(1) \Lambda^{J ]}(2)
 \Lambda^{I_{1}}(1) \Lambda^{I_{2}}(1) \Lambda^{I_{3}}(2) \Lambda^{I_{4}}(2) \right\}
  e^{i x^{\mu} P_{\mu}} | \tilde{0} \rangle \nn \\
& & - \frac{1}{4} \theta_{\alpha I} \theta_{\beta J} \left\{ \overline{s}^{( \alpha}(1)
s^{\beta )}(1) \left( \Omega^{I J} \Lambda^{I_{1}}(1) \Lambda^{I_{2}}(1) \Lambda^{I_{3}}(2)
\Lambda^{I_{4}}(2) + 4 \Lambda^{[ I}(1) \Omega^{J ] [I_{1}} \Lambda^{I_{2} ]}(1)
 \Lambda^{I_{3}}(2) \Lambda^{I_{4}}(2) \right) \right. \nn \\
& & \hspace{1.8cm} + \overline{s}^{( \alpha}(2) s^{\beta )}(2) \left( \Omega^{I J}
\Lambda^{I_{1}}(1) \Lambda^{I_{2}}(1) \Lambda^{I_{3}}(2) \Lambda^{I_{4}}(2) + 4
\Lambda^{[ I}(2) \Omega^{J ] [I_{3}} \Lambda^{I_{4} ]}(2) \Lambda^{I_{1}}(1)
 \Lambda^{I_{2}}(1) \right) \nn \\
& & \hspace{1.8cm} + 4 \overline{s}^{( \alpha}(1) s^{\beta )}(2) \Lambda^{[ I}(2)
 \Omega^{J ] [I_{1}} \Lambda^{I_{2} ]}(1) \Lambda^{I_{3}}(2) \Lambda^{I_{4}}(2) \nn \\
& & \hspace{1.8cm} \left. + 4 \overline{s}^{( \alpha}(2) s^{\beta )}(1) \Lambda^{[ I}(1)
\Omega^{J ] [I_{3}} \Lambda^{I_{4} ]}(2) \Lambda^{I_{1}}(1) \Lambda^{I_{2}}(1) \right\}
e^{i x^{\mu} P_{\mu}} | \tilde{0} \rangle \nn \\
& & - 2 \theta_{\alpha I} \theta_{\beta J} \overline{s}^{[ \alpha}(1) \overline{s}^{\beta ]}(2)
 \Lambda^{[ I_{1}}(1) \Omega^{I_{2} ] ( I} \Omega^{J ) [ I_{3}} \Lambda^{I_{4} ]}(2) e^{i x^{\mu}
 P_{\mu}} | \tilde{0} \rangle \nn \\
& & - \frac{1}{2} \theta_{\alpha I} \theta_{\beta J} s^{[ \alpha}(1) s^{\beta ]}(2)
 \Lambda^{( I}(1) \Lambda^{J )}(2) \Lambda^{I_{1}}(1) \Lambda^{I_{2}}(1) \Lambda^{I_{3}}(2)
  \Lambda^{I_{4}}(2) e^{i x^{\mu} P_{\mu}} | \tilde{0} \rangle \nn \\
& & + \frac{1}{2} \theta_{\alpha I} \theta_{\beta J} \left\{ \overline{s}^{[ \alpha}(1)
s^{\beta ]}(2) \Lambda^{( I}(2) \Omega^{J ) [ I_{1}} \Lambda^{I_{2} ]}(1) \Lambda^{I_{3}}(2)
\Lambda^{I_{4}}(2) \right. \nn \\
& & \hspace{1.8cm} \left. + \overline{s}^{[ \alpha}(2) s^{\beta ]}(1) \Lambda^{( I}(1)
\Omega^{J ) [ I_{3}} \Lambda^{I_{4} ]}(2) \Lambda^{I_{1}}(1) \Lambda^{I_{2}}(1)
\right\} e^{i x^{\mu} P_{\mu}} | \tilde{0} \rangle \nn \\
& & - \frac{1}{\sqrt{2}} \theta_{\alpha I} \theta_{\beta J} \theta_{\gamma K}
 \left\{ \overline{s}^{\alpha}(1) \overline{s}^{[ \beta}(1) \overline{s}^{\gamma ]}(2)
  \Omega^{I [ I_{1}} \Omega^{I_{2} ] ( J} \Omega^{K ) [ I_{3}} \Lambda^{I_{4} ]}(2) \right.
   \nn \\
& & \hspace{2.8cm} \left. + \overline{s}^{\alpha}(2) \overline{s}^{[ \beta}(2)
 \overline{s}^{\gamma ]}(1) \Omega^{I [ I_{3}} \Omega^{I_{4} ] ( J} \Omega^{K )
 [ I_{1}} \Lambda^{I_{2} ]}(1) \right\} e^{i x^{\mu} P_{\mu}} | \tilde{0} \rangle \nn \\
& & - \frac{1}{4 \sqrt{2}} \theta_{\alpha I} \theta_{\beta J} \theta_{\gamma K}
 \left\{ s^{\alpha}(1) s^{[ \beta}(1) s^{\gamma ]}(2) \Lambda^{I}(1) \Lambda^{( J}(1)
 \Lambda^{K )}(2) \Lambda^{I_{1}}(1) \Lambda^{I_{2}}(1) \Lambda^{I_{3}}(2) \Lambda^{I_{4}}(2)
 \right. \nn \\
& & \hspace{2.8cm} \left. + s^{\alpha}(2) s^{[ \beta}(2) s^{\gamma ]}(1) \Lambda^{I}(2)
\Lambda^{( J}(2) \Lambda^{K )}(1) \Lambda^{I_{1}}(1) \Lambda^{I_{2}}(1) \Lambda^{I_{3}}(2)
 \Lambda^{I_{4}}(2) \right\} e^{i x^{\mu} P_{\mu}} | \tilde{0} \rangle \nn \\
& & - \frac{1}{2 \sqrt{2}} \theta_{\alpha I} \theta_{\beta J} \theta_{\gamma K} \left\{
\overline{s}^{\alpha}(1) \overline{s}^{[ \beta}(1) s^{\gamma ]}(2) \Omega^{I [ I_{1}}
\Omega^{I_{2} ] ( J} \Lambda^{K )}(2) \Lambda^{I_{3}}(2) \Lambda^{I_{4}}(2) \right. \nn \\
& & \hspace{2.8cm} \left. + \overline{s}^{\alpha}(2) \overline{s}^{[ \beta}(2) s^{\gamma ]}(1)
\Omega^{I [ I_{3}} \Omega^{I_{4} ] ( J} \Lambda^{K )}(1) \Lambda^{I_{1}}(1) \Lambda^{I_{2}}(1)
 \right\} e^{i x^{\mu} P_{\mu}} | \tilde{0} \rangle \nn \\
& & - \frac{1}{2 \sqrt{2}} \theta_{\alpha I} \theta_{\beta J} \theta_{\gamma K} \left\{
 \overline{s}^{[ \alpha}(1) s^{\beta ]}(2) s^{\gamma}(2) \Lambda^{[ I_{1}}(1) \Omega^{I_{2} ]
 ( I} \Lambda^{J )}(2) \Lambda^{K}(2) \Lambda^{I_{3}}(2) \Lambda^{I_{4}}(2) \right. \nn \\
& & \hspace{2.8cm} \left. + \overline{s}^{[ \alpha}(2) s^{\beta ]}(1) s^{\gamma}(1)
\Lambda^{[ I_{3}}(2) \Omega^{I_{4} ] ( I} \Lambda^{J )}(1) \Lambda^{K}(1) \Lambda^{I_{1}}(1)
 \Lambda^{I_{2}}(1) \right\} e^{i x^{\mu} P_{\mu}} | \tilde{0} \rangle \nn \\
& & - \frac{1}{4} \theta_{\alpha I} \theta_{\beta J} \theta_{\gamma K} \theta_{\delta L}
\overline{s}^{[ \alpha}(1) \overline{s}^{\beta ]}(2) \overline{s}^{[ \gamma}(1)
 \overline{s}^{\delta ]}(2) \Omega^{I_{1} ( I} \Omega^{J ) I_{3}} \Omega^{I_{2}
 ( K} \Omega^{L ) I_{4}} e^{i x^{\mu} P_{\mu}} | \tilde{0} \rangle \nn \\
& & + \frac{1}{16} \theta_{\alpha I} \theta_{\beta J} \theta_{\gamma K} \theta_{\delta L}
 s^{[ \alpha}(1) s^{\beta ]}(2) s^{[ \gamma}(1) s^{\delta ]}(2) \Lambda^{( I}(1) \Lambda^{J )}(2)
  \Lambda^{( K}(1) \Lambda^{L )}(2) \Lambda^{I_{1}}(1) \Lambda^{I_{2}}(1) \Lambda^{I_{3}}(2)
  \Lambda^{I_{4}}(2) e^{i x^{\mu} P_{\mu}} | \tilde{0} \rangle \nn \\
& & - \frac{1}{8} \theta_{\alpha I} \theta_{\beta J} \theta_{\gamma K} \theta_{\delta L}
\left\{ \overline{s}^{[ \alpha}(1) s^{\beta ]}(2) \overline{s}^{[ \gamma}(1) s^{\delta ]}(2)
 \Lambda^{( I}(2) \Omega^{J ) [ I_{1}} \Omega^{I_{2} ] ( K} \Lambda^{L )}(2) \Lambda^{I_{3}}(2)
  \Lambda^{I_{4}}(2) \right. \nn \\
& & \hspace{3cm} \left. + \overline{s}^{[ \alpha}(2) s^{\beta ]}(1)
\overline{s}^{[ \gamma}(2) s^{\delta ]}(1) \Lambda^{( I}(1) \Omega^{J ) [ I_{3}}
\Omega^{I_{4} ] ( K} \Lambda^{L )}(1) \Lambda^{I_{1}}(1) \Lambda^{I_{2}}(1) \right\}
 e^{i x^{\mu} P_{\mu}} | \tilde{0} \rangle \nn \\
& & + \mbox{derivative terms} \nn \\
& & + \mathcal{O} ( \theta \theta \theta \theta \theta ) \nn \\
& \cong & \left[ \mbox{(0,0,0)}_{D},\underline{14} \right] + \left[ \mbox{(1,0,0)}_{D},
\underline{16} \right] + \left[ \mbox{(2,0,0)}_{D},\underline{5} \right] +
\left[ \mbox{(0,1,0)}_{D},\underline{10} \right] + \left[ \mbox{(1,1,0)}_{D},\underline{4} \right]
 + \left[ \mbox{(0,2,0)}_{D},\underline{1} \right] \nn \\
& & + \mbox{derivative terms.}
\enn

It is important to note that, the terms $\left( \overline{s}^{( \alpha}(1)
 \overline{s}^{\beta )}(2) \ldots \right)$, $\left( s^{( \alpha}(1) s^{\beta )}(2) \ldots
  \right)$, $\left( \overline{s}^{( \alpha}(1) s^{\beta )}(2) \ldots \right)$ and $\left(
  \overline{s}^{( \alpha}(2) s^{\beta )}(1) \ldots \right)$ on the right hand side above as
   well as the terms of the form $\left( \overline{s}^{( \alpha} \overline{s}^{\beta}
   \overline{s}^{\gamma )} \ldots \right)$, $\left( \overline{s}^{[ \alpha}
   \overline{s}^{( \beta ]} \overline{s}^{\gamma} \overline{s}^{\delta )}
   \ldots \right)$ and $\left( \overline{s}^{( \alpha} \overline{s}^{\beta}
   \overline{s}^{\gamma} \overline{s}^{\delta )} \ldots \right)$ (plus those
   others which are obtained by replacing one or more $\overline{s}$ with corresponding $s$)
    vanish when acted upon by $\left( \mathcal{P}_{X Y} \right)_{I_{1} I_{2} I_{3} I_{4}}$.

In the table below, we give only one term from each type which corresponds to a different
lowest weight vector of the graviton supermultiplet. \vspace{.2cm}
\begin{center}
\begin{tabular}{|c|c|c|}
\hline
$SU(4) \times SU(2)$ & $SU^{*}(4) \times USp(4)$ & $USp(4)$ \\ \hline \hline
$| 1 , 1 \rangle$ & $\left( \mathcal{P}_{X Y} \right)_{I_{1} I_{2} I_{3} I_{4}}
 \Lambda^{I_{1}}(1) \Lambda^{I_{2}}(1) \Lambda^{I_{3}}(2) \Lambda^{I_{4}}(2)
  e^{i x^{\mu} P_{\mu}} | \tilde{0} \rangle$ & 14 \\ \hline
$| \onebox , \onebox \rangle$ & $\overline{s}^{\alpha}(1)
\left( \mathcal{P}_{X Y} \right)_{I_{1} I_{2} I_{3} I_{4}} \Omega^{I [ I_{1}}
 \Lambda^{I_{2} ]}(1) \Lambda^{I_{3}}(2) \Lambda^{I_{4}}(2) e^{i x^{\mu}
 P_{\mu}} | \tilde{0} \rangle $ & 16 \\ \hline
$|\twobox, \oneonebox \rangle$ & $\overline{s}^{( \alpha}(1) \overline{s}^{\beta )}(1)
 \left( \mathcal{P}_{X Y} \right)_{I_{1} I_{2} I_{3} I_{4}}
  \Omega^{I [ I_{1}} \Omega^{I_{2} ] J} \Lambda^{I_{3}}(2) \Lambda^{I_{4}}(2) e^{i x^{\mu}
   P_{\mu}} | \tilde{0} \rangle$ & 5 \\ \hline
$|\oneonebox, \twobox \rangle$ & $\overline{s}^{[ \alpha}(1) \overline{s}^{\beta ]}(2)
\left( \mathcal{P}_{X Y} \right)_{I_{1} I_{2} I_{3} I_{4}} \Lambda^{[ I_{1}}(1) \Omega^{I_{2} ]
 ( I} \Omega^{J ) [ I_{3}} \Lambda^{I_{4} ]}(2) e^{i x^{\mu} P_{\mu}} | \tilde{0} \rangle$ & 10
 \\ \hline
$|\twoonebox, \twoonebox \rangle$ & $\overline{s}^{\alpha}(1) \overline{s}^{[ \beta}(1)
 \overline{s}^{\gamma ]}(2) \left( \mathcal{P}_{X Y} \right)_{I_{1} I_{2} I_{3} I_{4}}
  \Omega^{I [ I_{1}} \Omega^{I_{2} ] ( J} \Omega^{K ) [ I_{3}} \Lambda^{I_{4} ]}(2)
   e^{i x^{\mu} P_{\mu}} | \tilde{0} \rangle$ & 4 \\ \hline
$|\twotwobox, \twotwobox \rangle$ & $\overline{s}^{[ \alpha}(1) \overline{s}^{\beta ]}(2)
\overline{s}^{[ \gamma}(1) \overline{s}^{\delta ]}(2) \left(
\mathcal{P}_{X Y} \right)_{I_{1} I_{2} I_{3} I_{4}} \Omega^{I_{1}
( I} \Omega^{J ) I_{3}} \Omega^{I_{2} ( K} \Omega^{L ) I_{4}} e^{i x^{\mu} P_{\mu}} |
\tilde{0} \rangle$ & 1 \\ \hline
\end{tabular}
\end{center}
\begin{center} Table 8. The massless graviton supermultiplet in the
 supercoherent state basis. \end{center}
\vspace{.2cm}

\section{Conclusion}
\setcounter{equation}{0}

In this paper, we have studied the positive energy unitary
 representations of $2N$ extended superconformal algebras in six
 dimensions in a supercoherent state basis. These supercoherent states
 represent  conformal superfields in $d=6$. The ultra short
 doubleton supermultiplets  of \OSN ~ correspond to  the
 massless conformal  superfields in $d = 6$. The massive conformal
 superfields  are defined by those representations of \OSN ~that are
 obtainable by  tensoring these doubleton supermultiplets. Those
 massive superfields  obtained by tensoring {\it two} copies of the
 doubletons  correspond to massless $AdS_{7}$ supermultiplets. For $N
 = 2$,  the CPT self-conjugate supermultiplet obtained by tensoring
 two copies  of CPT self-conjugate doubleton supermultiplets is simply
 the graviton  supermultiplet of  $AdS$ supergroup \OS. We give explicitly
 the  supercoherent state basis of the graviton supermultiplet and the
 corresponding superfield. Even though the doubleton representations
 of \OSN~  do not have a Poincar\'{e} limit the representations obtained
by tensoring doubletons with each other have a Poincar\'{e} limit
in $d=7$.

The supermultiplets that are ``shortened''(i.e short or
of intermediate length)  correspond to BPS supermultiplets  preserving various
 amounts of supersymmetry. Using the results of this paper one can
 study BPS supermultiplets in a supercoherent state basis.

 Our results can also be used to write down explicitly infinite spin
 anti-de  Sitter superalgebras as suggested in \cite{mg89}. These
 infinite spin  superalgebras have been studied by many authors, in
 particular  by M.A.~Vasiliev, et.al. \cite{vasiliev}. The oscillator
 realization of $N = 8$  $AdS_{5}$ superalgebra $SU(2,2|4)$ \cite{mgnm,gmz2}, was used
 recently  to write down infinite spin superalgebras in $d = 5$
 \cite{sezgin}.  One can use the  realization of \OSN ~ in the
 non-compact basis to  write down  infinite spin $AdS_7$ superalgebras
  in a covariant basis and study its unitary representations.

Finally, we should stress that one can define supercoherent state
 bases for all  non-compact supergroups that admit positive energy
 unitary representations \cite{mg00}.

\section*{Acknowledgements}
SF was supported in part by a Braddock Fellowship.

\section*{Appendix}
\appendix
\section{Notations}
\def\theequation{A.\arabic{equation}}
\setcounter{equation}0
Here we give a list of indices we used in this paper and their ranges :
\eqn
\mu,\nu,\rho,\sigma = 0,1,\ldots,5 & \hspace{1cm} & \mbox{six dimensional Minkowski
spacetime indices} \nn \\
a,b,c,d = 0,1,\ldots,7 & \hspace{1cm} & \mbox{$SO^{*}(8)$ ($\cong SO(6,2)$) vector indices} \nn
\\
m,n = 1,2,\ldots,6 & \hspace{1cm} & \mbox{$SO(6)$ vector indices} \nn \\
\hat{m},\hat{n} = 1,2,\ldots,6 & \hspace{1cm} & \mbox{$SO_{c}(5,1)$ vector indices} \nn \\
i,j,k,l = 1,2,3,4 & \hspace{1cm} & \mbox{$SU(4)$ indices} \nn \\
\hat{i},\hat{j},\hat{k},\hat{l} = 1,2,3,4 & \hspace{1cm} & \mbox{$SU_{c}^{*}(4)$ spinor  indices}
 \nn \\
A,B,C,D = 1,2,\ldots,8 & \hspace{1cm} & \mbox{$SO^{*}(8)$  ($\cong SO(6,2)$) left-handed
spinor indices} \nn \\
I,J,K,L = 1,2,\ldots,2N & \hspace{1cm} & \mbox{$USp(2N)$ indices in the fundamental
representation} \nn \\
\kappa,\lambda, = 1,2,\ldots,N & \hspace{1cm} & \mbox{$SU(N)$ indices in the fundamental
 representation} \nn \\
M,N = 1,2,3,4|1,\ldots,N & \hspace{1cm} & \mbox{$SU(4|N)$ indices in the fundamental
representation} \nn \\
\alpha,\beta,\gamma,\delta = 1,2,3,4 & \hspace{1cm} & \mbox{$SU^{*}(4)$ left-handed  spinor
indices} \nn \\
\dot{\alpha},\dot{\beta},\dot{\gamma},\dot{\delta} = 1,2,3,4 & \hspace{1cm}
& \mbox{$SU^{*}(4)$ right-handed spinor indices} \nn \\
X,Y,Z = 1,2,\ldots,5 & \hspace{1cm} & \mbox{Internal $SO(5) (\cong USp(4))$ vector indices}
\nn \\
r,s,t,u = 1,2,\ldots,P & \hspace{1cm} & \mbox{color indices} \nn
\enn
\vspace{.5cm}

The $USp(2N)$ symplectic invariant tensor is taken to be
\eq
\Omega_{IJ} = \Omega^{IJ} = \left( \begin{array}{cc} O & I \\ - I & O \end{array}
 \right)_{2N \times 2N},
\en
so that
\eq
\Omega_{I J} \Omega^{J K} = - \delta_{I}^{~K},
\en
and is used to lower/raise the $USp(2N)$ spinor indices as follows :
\eq
\lambda_{I} = \lambda^{J} \Omega_{JI}, \quad \lambda^{I} = \Omega^{IJ} \lambda_{J}.
\en
\section{Six dimensional $\Gamma$ matrices}
\def\theequation{B.\arabic{equation}}
\setcounter{equation}0

Our choice of $\Gamma$-matrices is given below:
\eqn
\Gamma_{0} & = & \sigma_{3} \otimes I_{2} \otimes I_{2}, \nn \\
\Gamma_{1} & = & i \sigma_{1} \otimes \sigma_{2} \otimes I_{2}, \nn \\
\Gamma_{2} & = & i \sigma_{1} \otimes \sigma_{1} \otimes \sigma_{2}, \nn \\
\Gamma_{3} & = & i \sigma_{1} \otimes \sigma_{3} \otimes \sigma_{2}, \nn \\
\Gamma_{4} & = & i \sigma_{2} \otimes I_{2} \otimes \sigma_{2} \nn, \\
\Gamma_{5} & = & i \sigma_{2} \otimes \sigma_{2} \otimes \sigma_{1}, \enn where
$\sigma_{1}$, $\sigma_{2}$ and $\sigma_{3}$ are the Pauli matrices and $I_{2}$ is  the $2
\times 2$ identity matrix.

Therefore, \eq \Gamma_{7} = - \Gamma_{0} \Gamma_{1} \Gamma_{2}
\Gamma_{3} \Gamma_{4} \Gamma_{5} = - \sigma_{2} \otimes \sigma_{2} \otimes \sigma_{3}.
\en

All these matrices have the index structure $\left( \Gamma_{\mu} \right)_{A}^{~B}$.
We raise or lower these $SO^{*}(8)$ left-handed spinor indices using the charge
conjugation matrix $C$ as shown below : \eqn
\left( \Gamma_{\mu} \right)_{A B} & = & \left( \Gamma_{\mu} \right)_{A}^{~C} C_{C B}, \nn \\
\left( \Gamma_{\mu} \right)^{A B} & = & C^{A C} \left( \Gamma_{\mu} \right)_{C}^{~B},
\enn
where
\eqn
C_{A B} = C^{A B} & = & \left( \begin{array}{cc} O & I \\ I & O \end{array}
\right)_{8 \times 8} \nn \\
& = & - \Gamma_{1} \Gamma_{2} \Gamma_{3} \nn \\
& = & i \Gamma_{0} \Gamma_{4} \Gamma_{5} \Gamma_{7}.
\enn
\section{$SU^{*}(4)$ $\Sigma$ matrices}
\def\theequation{C.\arabic{equation}}
\setcounter{equation}0

The $\Sigma$ matrices in $d = 6$, the analogs of Pauli matrices
$\sigma_{\mu}$ in $d = 4$,  which satisfy the equation (\ref{pksigma}) are given below.
\eqn
\begin{array}{rclcrcl}
\left( \Sigma_{0} \right)_{\alpha}^{~\dot{\beta}} & = & - i \sigma_{2} \otimes \sigma_{3},
 & \hspace{3cm} & \left( \Sigma_{0} \right)_{\dot{\alpha}}^{~\beta} & = & i \sigma_{2}
  \otimes \sigma_{3}, \\
\left( \Sigma_{1} \right)_{\alpha}^{~\dot{\beta}} & = & i \sigma_{2} \otimes I_{2}, &
 \hspace{3cm} & \left( \Sigma_{1} \right)_{\dot{\alpha}}^{~\beta} & = & - i \sigma_{2}
 \otimes I_{2}, \\
\left( \Sigma_{2} \right)_{\alpha}^{~\dot{\beta}} & = & i \sigma_{1} \otimes \sigma_{2},
 & \hspace{3cm} & \left( \Sigma_{2} \right)_{\dot{\alpha}}^{~\beta} & = & - i \sigma_{1}
 \otimes \sigma_{2}, \\
\left( \Sigma_{3} \right)_{\alpha}^{~\dot{\beta}} & = & i \sigma_{3} \otimes \sigma_{2},
& \hspace{3cm} & \left( \Sigma_{3} \right)_{\dot{\alpha}}^{~\beta} & = & - i \sigma_{3}
 \otimes \sigma_{2}, \\
\left( \Sigma_{4} \right)_{\alpha}^{~\dot{\beta}} & = & I_{2} \otimes \sigma_{2}, &
\hspace{3cm} & \left( \Sigma_{4} \right)_{\dot{\alpha}}^{~\beta} & = & I_{2} \otimes
\sigma_{2}, \\
\left( \Sigma_{5} \right)_{\alpha}^{~\dot{\beta}} & = & \sigma_{2} \otimes \sigma_{1},
& \hspace{3cm} & \left( \Sigma_{5} \right)_{\dot{\alpha}}^{~\beta} & = & \sigma_{2} \otimes
\sigma_{1}.
\end{array}
\enn

Equations (\ref{sigmasecondset}) lead one to the following matrices :
\eqn
\begin{array}{rclcrcl}
\left( \Sigma_{0} \right)_{\alpha}^{~\beta} & = & - I_{2} \otimes I_{2}, & \hspace{3cm} &
\left( \Sigma_{0} \right)_{\dot{\alpha}}^{~\dot{\beta}} & = & - I_{2} \otimes I_{2}, \\
\left( \Sigma_{1} \right)_{\alpha}^{~\beta} & = & I_{2} \otimes \sigma_{3}, & \hspace{3cm} &
 \left( \Sigma_{1} \right)_{\dot{\alpha}}^{~\dot{\beta}} & = & I_{2} \otimes \sigma_{3}, \\
\left( \Sigma_{2} \right)_{\alpha}^{~\beta} & = & - \sigma_{3} \otimes \sigma_{1}, & \hspace{3cm}
 & \left( \Sigma_{2} \right)_{\dot{\alpha}}^{~\dot{\beta}} & = & - \sigma_{3} \otimes \sigma_{1},
  \\
\left( \Sigma_{3} \right)_{\alpha}^{~\beta} & = & \sigma_{1} \otimes \sigma_{1}, & \hspace{3cm}
& \left( \Sigma_{3} \right)_{\dot{\alpha}}^{~\dot{\beta}} & = & \sigma_{1} \otimes \sigma_{1}, \\
\left( \Sigma_{4} \right)_{\alpha}^{~\beta} & = & \sigma_{2} \otimes \sigma_{1}, & \hspace{3cm} &
\left( \Sigma_{4} \right)_{\dot{\alpha}}^{~\dot{\beta}} & = & - \sigma_{2} \otimes \sigma_{1}, \\
\left( \Sigma_{5} \right)_{\alpha}^{~\beta} & = & - I_{2} \otimes \sigma_{2}, & \hspace{3cm} &
 \left( \Sigma_{5} \right)_{\dot{\alpha}}^{~\dot{\beta}} & = & I_{2} \otimes \sigma_{2}.
\end{array}
\enn
\section{$SO(5)$ $\gamma$ matrices}
\def\theequation{D.\arabic{equation}}
\setcounter{equation}0

In order for one to express the lowest weight vectors in the massless (in $AdS_{7}$ sense)
 graviton supermultiplet in a $USp(4)$ covariant form, we introduce the $4 \times 4$ gamma
 matrices of $Spin(5) ( \cong SO(5) )$ $( \gamma_{X} )_{I}^{~J}$ where $X = 1,2,\ldots,5$
  are the vector indices and $I,J = 1,2,3,4$ are the spinor indices.

These gamma matrices satisfy the Clifford algebra
\eq
\{ \gamma_{X}, \gamma_{Y} \} = 2 \delta_{XY},
\en
and are symplectic traceless :
\eq
\Omega^{IJ} \left( \gamma_{X} \right)_{IJ} \hspace{0.2cm} = \hspace{0.2cm}
\left( \gamma_{X} \right)_{I}^{~I} \hspace{0.2cm} = \hspace{0.2cm} 0.
\en

The matrices
\eq
\Sigma ( M_{X Y} ) = \frac{i}{4} [ \gamma_{X} , \gamma_{Y} ]
\en
generate a four dimensional spinor representation of the algebra of $Spin(5)$,
\eq
[ M_{X Y} , M_{X^{\prime} Y^{\prime}} ] = i ( \delta_{Y X^{\prime}} M_{X Y^{\prime}} -
\delta_{X X^{\prime}} M_{Y Y^{\prime}} - \delta_{Y Y^{\prime}} M_{X X^{\prime}} -
\delta_{X Y^{\prime}} M_{Y X^{\prime}} ).
\en

It is worth noting that the $USp(4)$ spinor indices of these $\gamma$-matrices are lowered
(raised) by use of the symplectic invariant tensor $\Omega_{IJ}$ ($\Omega^{IJ}$) introduced
 in the paper. In particular, we have
\eqn
( \gamma_{X} )^{IJ} & = & \Omega^{IK} \Omega^{JL} ( \gamma_{X} )_{KL}, \nn \\
( \gamma_{X} )_{IJ} & = & ( \gamma_{X} )^{KL} \Omega_{KI} \Omega_{LJ}.
\enn

Moreover, these gamma matrices can be chosen so that the type (1,1) $( \gamma_{X} )_{I}^{~J}$
form are pure anti-hermitian and the type (0,2) $( \gamma_{x} )_{IJ}$ and type (2,0)
$( \gamma_{x} )^{IJ}$ forms are anti-symmetric, viz.,
\eq
( \gamma_{x} )_{I}^{~J} \Omega_{JK} \hspace{0.2cm} = \hspace{0.2cm} ( \gamma_{x} )_{IK}
 \hspace{0.2cm} = \hspace{0.2cm} - ( \gamma_{x} )_{KI},
\en
and
\eq
( ( \gamma_{x} )_{I}^{~J} )^{\dag} \hspace{0.2cm} = \hspace{0.2cm} ( \gamma_{x}^{*} )_{J}^{~I}
 \hspace{0.2cm} = \hspace{0.2cm} - ( \gamma_{x} )_{I}^{~J}.
\en

Now from \cite{mgst}, we know that the lowest weight vectors in the massless graviton
 supermultiplet transform under one of the following irreducible representations of $USp(4)$;
 1, 4, 5, 10, 14 or 16. We consider each of the states in the massless graviton supermultiplet
 separately. We let the color indices $r,s = 1,2$.
\begin{itemize}
\item $| \tilde{0} \rangle$ is a $\underline{1}$ of $USp(4)$.
\item $\Lambda^{I}(r) | \tilde{0} \rangle$ is a $\underline{4}$ of $USp(4)$.
\item $\Lambda^{I}(r) \Lambda^{J}(s) | \tilde{0} \rangle$ is reducible under $USp(4)$. One
 finds the following decomposition into irreducible components :
\eqn
\Lambda^{I}(r) \Lambda^{J}(s) | \tilde{0} \rangle & \rightarrow & \Lambda^{( I}(r)
 \Lambda^{J )}(s) \epsilon_{rs} | \tilde{0} \rangle  \oplus ( \gamma_{X} )_{IJ} \Lambda^{[ I}(r)
  \Lambda^{J ]}(s) | \tilde{0} \rangle \nn \\
& & \oplus \Lambda^{I}(r) \Lambda^{J}(s) \Omega_{IJ} | \tilde{0} \rangle \nn \\
& = & \underline{10} \oplus \underline{5} \oplus \underline{1}.
\enn
\item $\Lambda^{I}(r) \Lambda^{J}(s) \Lambda^{K}(t) | \tilde{0} \rangle$ is reducible.
It can be decomposed into irreducible parts as follows :
\eqn
\Lambda^{I}(r) \Lambda^{J}(s) \Lambda^{K}(t) | \tilde{0} \rangle & \rightarrow &
\Lambda^{[ I}(r) \Lambda^{J}(s) \Lambda^{K ]}(t) | \tilde{0}
\rangle \oplus \tilde{\psi}_{X}^{~K} | \tilde{0} \rangle \nn \\
& = & \underline{4} \oplus \underline{16}.
\enn
where we have defined
\eqn
\psi_{X}^{~K} & = & ( \gamma_{X} )_{IJ} \Lambda^{I}(r) \Lambda^{J}(s) \Lambda^{K}(t) \nn \\
\tilde{\psi}_{X}^{~K} & = & \psi_{X}^{~K} -
\frac{1}{5} ( \gamma_{X} )_{J}^{~K} \left[ ( \gamma^{Y} )^{J}_{~I}\psi_{Y}^{~I} \right].
\enn
\end{itemize}

\end{document}